\documentclass[a4paper,10pt]{article}
\pdfoutput=1
\usepackage{jheppub}

\usepackage{graphicx,wrapfig,float,slashed,cancel,bm}
\usepackage{amsmath,amssymb,epsfig,graphicx,xcolor,stmaryrd}
\usepackage{mathtools}
\usepackage{epstopdf}
\usepackage{soul}
\usepackage{ragged2e}
\usepackage{lipsum}
\usepackage{colortbl}
\usepackage{tabu}
\usepackage{physics}
\usepackage{verbatim}
\allowdisplaybreaks
\usepackage{pifont}
\usepackage[normalem]{ulem}
\usepackage{microtype}
\usepackage{subfig}
\usepackage{titlesec}
\usepackage[font={it},skip=0pt]{caption}	

\newcommand{\nc}{\newcommand}
\nc{\non}{\nonumber}
\nc{\hc}{\hbox {h.c.}}
\nc{\noi}{\noindent}
\nc{\barx}{\bar{x}}

\nc{\hsp}{\hspace{0.5cm}}
\nc{\lsp}{\hspace{1cm}}
\nc{\Lsp}{\hspace{2cm}}
\nc{\LLsp}{\lsp\lsp}
\nc{\lra}{\longrightarrow}
\nc{\p}{\prime}
\nc{\sgn}{\text{sgn}}
\nc{\ph}{\varphi}
\nc{\eq}{\text{Eq.~}}
\nc{\cL}{\mathcal{L}}
\nc{\vmol}{v_{\text{M\o l}}}
\nc{\eff}{\text{eff}}
\nc{\sm}{\text{\rm SM}}

\nc{\dphi}{\partial_\phi}
\nc{\DLL}{\Delta_{LL}^q(\phi,\phi';-p^2)}
\nc{\DRL}{\Delta_{RL}^q(\phi,\phi';-p^2)}
\nc{\DRR}{\Delta_{RR}^q(\phi,\phi';-p^2)}
\nc{\DLR}{\Delta_{LR}^q(\phi,\phi';-p^2)}
\nc{\sgnphi}{{\rm sgn} (\phi)}
\nc{\ome}{\omega (\phi)}
\nc{\Ss}{S(x,\phi)}
\nc{\siphi}{\sigma(\phi)}
\nc{\noin}{\noindent}
\nc{\vp}{\varphi}
\nc{\tkappa}{\tilde{\kappa}}
\nc{\tlambda}{\tilde{\lambda}}
\nc{\tH}{\tilde{H}}
\nc{\DepsK}{\Delta \epsilon_K}
\nc{\epsK}{\epsilon_K}

\DeclareMathOperator{\atanh}{arctanh}

\def\lb{\Lambda_{\textsc b}}

\def\e2a{e^{2A(y)}}

\def\eq{Eq.~}
\def\ms{M_\ast}
\def\mpl{M_{\text{Pl}}}

\def\mkk{M_{\rm KK}}
\nc{\beq}{\begin{equation}}  \nc{\eeq}{\end{equation}}
\nc{\bea}{\begin{eqnarray}}  \nc{\eea}{\end{eqnarray}}
\nc{\baa}{\begin{array}}     \nc{\eaa}{\end{array}}
\nc{\bit}{\begin{itemize}}   \nc{\eit}{\end{itemize}}
\nc{\ben}{\begin{enumerate}} \nc{\een}{\end{enumerate}}
\nc{\bce}{\begin{center}}    \nc{\ece}{\end{center}}
\nc{\bpm}{\begin{pmatrix}}   \nc{\epm}{\end{pmatrix}}
\nc{\bvt}{\begin{verbatim}}  \nc{\evt}{\end{verbatim}}

\def\lsim{\mathrel{\raise.3ex\hbox{$<$\kern-.75em\lower1ex\hbox{$\sim$}}}}
\def\gsim{\mathrel{\raise.3ex\hbox{$>$\kern-.75em\lower1ex\hbox{$\sim$}}}}

\def\udots{\mathinner{\mkern1mu\raise1pt\vbox{\kern7pt\hbox{.}}\mkern2mu\raise4pt\hbox{.}\mkern2mu\raise7pt\hbox{.}\mkern1mu}}

\def\tev{\;\hbox{\rm TeV}}

\begin{document}
\title{Dynamical origin of flavor hierarchies in a warped extra dimension}

\author[1,2]{Aqeel Ahmed,}
\author[1]{Adrian Carmona,}
\author[1]{Javier Castellano Ruiz,}
\author[1,3,4]{Yi Chung}
\author[1,5]{and Matthias Neubert}

\affiliation[1]{PRISMA$^{+}$ Cluster of Excellence \rm{\&} Mainz Institute for Theoretical Physics,\\ Johannes Gutenberg University, 55099 Mainz, Germany}
\affiliation[2]{Theoretische Natuurkunde \rm{\&} IIHE/ELEM,\\ Vrije Universiteit Brussel, Pleinlaan 2, 1050 Brussels, Belgium}
\affiliation[3]{Department of Physics, University of California Davis, Davis, California 95616, U.S.A.}
\affiliation[4]{Department of Physics, National Taiwan University, Taipei 10617, Taiwan}
\affiliation[5]{Department of Physics \rm{\&} LEPP, Cornell University, Ithaca, NY 14853, U.S.A.}

\emailAdd{aqeahmed@uni-mainz.de}\emailAdd{adcarmon@uni-mainz.de}\emailAdd{castella@uni-mainz.de}\emailAdd{yichung@ucdavis.edu}\emailAdd{matthias.neubert@uni-mainz.de}

\abstract{Extensions of the Standard Model featuring a warped extra dimension compactified on an $S^1/\mathbb{Z}_2$ orbifold, in which the fermions and gauge bosons live in the bulk of the fifth dimension, offer one of the most compelling mechanisms for addressing both the hierarchy problem and the flavor puzzle of the Standard Model. However, the five-dimensional mass terms of bulk fermions must be odd functions on the orbifold, and as such they should be described by a field depending on the coordinate of the extra dimension. We demonstrate the feasibility of dynamically generating these fermion bulk masses with a bulk scalar field in warped extra dimensions. The bulk scalar acquires a vacuum expectation value, which is odd under the orbifold symmetry and gives rise to the fermion bulk masses through non-universal Yukawa-like interactions. Like in the conventional Randall-Sundrum setup, the localization of the different fermion zero modes along the extra dimension naturally explains the observed flavor structure and four-dimensional mass hierarchy of the SM fermions. We study the phenomenological implications of the backreaction on the metric and the modified fermion profiles due to the bulk scalar field on electroweak precision and flavor observables. Using up-to-date data, we show that the contributions to the $S$, $T$, and $\epsilon_K$ parameters require the mass of the first Kaluza-Klein gluon resonance to be of order 14 and 10 TeV in the minimal and the custodial model, respectively, regardless of the effect of the backreaction.  Furthermore, effective flavor-changing interactions among the SM fermions induced by the bulk scalar are discussed. We also comment on  the potential impact of the Higgs portal interaction of the bulk scalar on the couplings of the Higgs boson.
}


\preprint{MITP/19-020}

\maketitle

\section{Introduction} \label{introduction}
The Standard Model (SM) of particle physics has been tested to an unprecedented precision at the Large Hadron Collider (LHC). In particular, the confirmation of the SM-like nature of the Higgs boson and the measurement of its interactions  at the LHC have provided an impressive example of the robustness of the SM as a coherent explanation of the laws of nature at very short distances. However, there are also several fundamental and open questions to which the SM cannot provide a successful answer, and this points to the presence of new phenomena. Besides the well-known \emph{hierarchy problem}, which has been one of the main motivations for extensions of the SM and has led to a multitude of interesting new theoretical frameworks, more mundane and pressing problems like e.g.\ the absence of a viable candidate for dark matter (DM), the unknown origin of fermion masses (\emph{the flavor puzzle}) or the observed baryon asymmetry of the universe require the presence of new dynamics beyond the SM (BSM). Even though it is not obvious that these problems should require such new phenomena to be at the reach of the LHC, we are confident that the SM needs to be completed and that one should left no stone unturned in the quest for new physics. In the landscape of new physics scenarios trying to address some of these questions, models with warped extra dimensions and their four-dimensional (4D) duals are among the most appealing ones since they provide a beautiful rationale to answer several of these questions (for a recent review on BSM scenarios see, e.g.\ \cite{Csaki:2016kln}). Randall-Sundrum (RS) models with one extra dimension \cite{Randall:1999ee} are a particular example which has drawn a lot of attention in the last two decades, since they can be seen as a toolbox for studying a more general class of models of Higgs and fermion compositeness~\cite{Maldacena:1997re, Gubser:1998bc, ArkaniHamed:2000ds, Rattazzi:2000hs, PerezVictoria:2001pa, Contino:2004vy, Gherghetta:2010cj}. The RS model introduces a warped extra dimension on an $S_1/\mathbb{Z}_2$ orbifold with two D3-branes at the fixed points of the orbifold. The patch of the extra dimension between these two branes features an anti-de Sitter (AdS) geometry. While gravity was allowed to propagate through the bulk of the extra dimension, the SM was originally confined to the D3 brane, on which the fundamental Planck scale of the theory is red-shifted to an effective infrared (IR) scale of order the TeV scale. Since the Higgs boson was localized with the rest of the SM in the IR brane, the hierarchy problem was solved because SM loops are naturally cut off at the scale of the IR brane. However, soon after the original RS proposal it was realized that a more natural setup was to allow the SM gauge and fermion fields to propagate in the extra dimension, such that the zero modes of these fields would correspond to the SM fields~\cite{Grossman:1999ra,Davoudiasl:1999tf,Gherghetta:2000qt,Huber:2000ie},  whereas the Higgs boson would still be localized on or near the IR brane. In particular, promoting the fermion fields to five-dimensional (5D) fields allows for an elegant explanation of the hierarchies seen in the SM fermion masses and mixings through the bulk fermion localization mechanism~\cite{Grossman:1999ra, Gherghetta:2000qt}. There have been an enormous body of work dedicated to the formal and phenomenological implications of the RS model and its extensions.

One important question arises if one examines the mechanism solving the flavor puzzle more closely. The bulk fermion localization mechanism relies on the fact that hierarchical Yukawa couplings can be generated via different localizations of the zero-mode profiles -- corresponding to the SM chiral fermions -- along the extra dimension, controlled by the values of their fermion bulk masses~\cite{Grossman:1999ra}. Then, similar values of the anarchic 5D Yukawa couplings can be compensated through exponentially different overlaps with the IR brane, where the Higgs field is localized, providing therefore a natural explanation of the SM fermion mass hierarchy. Importantly, however, the 5D bulk masses must be $\mathbb{Z}_2$-odd functions on the orbifold, which implies that they are {\em fields\/} on the extra dimension. The reason is that the 5D Dirac fermion bilinear $\bar\Psi_i \Psi_i$ is odd under the orbifold $\mathbb{Z}_2$ symmetry and therefore a constant mass term is forbidden by the orbifold symmetry. In existing implementations of this framework one introduces these $\mathbb{Z}_2$-odd bulk masses by hand, simply by multiplying constant mass parameters with a $\text{sgn}(\phi)$ function of the 5D coordinate $\phi\in[-\pi,\pi]$. Here we point out that in any consistent field-theoretic implementation of the fermion localization mechanism one should obtain the coordinate-dependent 5D mass parameters from the vacuum expectation value (VEV) of a $\mathbb{Z}_2$-odd bulk scalar field. The main goal of this work is to dynamically generate such a VEV and couple it to the bulk fermions in such a way that the corresponding zero modes can reproduce the observed SM fermions masses with $\mathcal{O}(1)$ parameters. To achieve this, we consider a bulk scalar field $\Sigma(x,\phi)$ which develops a non-trivial $\phi$-dependent VEV, where $x$ and $\phi$ denote the 4D and extra-dimensional coordinates, respectively. The scalar field $\Sigma(x,\phi)$ couples to 5D Dirac fermions through Yukawa interactions such that when the scalar acquires a VEV, $\langle \Sigma(x,\phi)\rangle\equiv \omega(\phi)$, the bulk fermion masses are generated dynamically, i.e.\ 
\begin{equation}
	\mathcal{Y}_i\Sigma(x,\phi)\bar\Psi_i(x,\phi)\Psi_i(x,\phi) \xrightarrow{\langle \Sigma(x,\phi)\rangle} M_i(\phi)\bar\Psi_i(x,\phi)\Psi_i(x,\phi),
\end{equation}
where $M_i(\phi)\equiv \mathcal{Y}_i\,\omega(\phi)$. We discuss the vacuum solutions of the odd bulk scalar in the flat and warped geometries. Since adding a scalar field with a non-trivial VEV may have effects on the metric, we also study its possible implications on the background geometry. A similar approach to generate bulk fermion masses has been discussed in the context of flat extra dimensions in~\cite{Georgi:00}. More generally, without any connection to generating bulk fermion masses, odd bulk scalars acquiring a non-trivial VEV were studied in the context of flat and warped extra dimensions in~\cite{Grzadkowski:2004mg,Toharia:2007xe,Toharia:2010ex}.
 
The phenomenological implications of our mechanism are important and lead to modifications w.r.t.\ the conventional RS model (with a sign-function in front of the fermion bulk masses) in the fermion zero-mode profiles of the SM fermions, the spectrum of Kaluza-Klein (KK) modes, and on the background geometry through backreaction on the metric (i.e., the warp factor). These effects are potentially important for deriving  constraints on the model from precision measurements, such as electroweak precision tests (EWPT)~\cite{Csaki:2002gy,Carena:2003fx,Carena:2004zn} and flavor observables~\cite{Agashe:2004cp, Csaki:2008zd, Blanke:2008zb, Bauer:2008xb, Bauer:2009cf, Bauer:2011ah}. We stress that the existence of a $\mathbb{Z}_2$-odd bulk scalar field should be seen as a generic feature of {\em any\/} realistic warped extra-dimension model with bulk matter fields. As a consequence, such models predict the existence of a new type of scalar KK states, whose masses are significantly larger than the masses of other KK resonances.

The paper is organized as follows: In section~\ref{sec:back}, we study the background solution of the odd bulk scalar field in flat and warped extra dimensions, while backreaction effects on the background geometry are discussed for the warped case. In section~\ref{sec:gauge}, we discuss the KK decomposition of the gauge sector assuming minimal (non-custodial) and custodial bulk gauge symmetries and study the implications on the EWPT due to the backreaction on the metric by the bulk scalar field. Furthermore, Appendix~\ref{app:gauge} contains the conventions and derivation of electroweak precision observables for the custodial case. The scalar sector is analyzed in section~\ref{sec:scalar} and fermion bulk masses generated from an odd bulk scalar are discussed in section~\ref{sec:fermions}. In particular, we examine the fermion zero modes and the corresponding KK modes of the fermion and scalar fields. The phenomenological implications of our model on the fermion mass hierarchies and mixings and on flavor observables are presented in section~\ref{sec:pheno}. In section~\ref{sec:Higgsportal}, the Higgs portal coupling to the bulk scalar is discussed. Finally, section~\ref{sec:conclusions} contains our summary and conclusions.

\section{Non-trivial background solution for an odd bulk scalar field}
\label{sec:back}

We consider a RS model with an odd bulk scalar in a warped extra-dimensional geometry. The extra dimension is defined on an $S_1/\mathbb{Z}_2$ orbifold with two D3-branes located at the fixed points of the orbifold, i.e.\ a ultraviolet (UV) brane at $\phi=0$ and an IR brane at $\phi=\pi$, where $\phi$ is the coordinate defining the position along the fifth-dimension. We define the 5D metric as
\begin{equation}
ds^2 = e^{-2\sigma(\phi)} \, \eta_{\mu\nu} dx^\mu dx^\nu - r^2 \, d \phi^2 , 	\label{metric}
\end{equation}
where $\sigma(\phi)$ is the warp factor, $\eta_{\mu\nu} ={\rm diag}(+1,-1,-1,-1)$ is the 4D Minkowski metric, $\phi\in[-\pi,\pi]$,  and $r$ is the compactification radius of the extra dimension. The metric ansatz~\eqref{metric} has the non-factorisable form and is the most general 5D metric which preserves 4D Poincar\'e invariance. In the case when the backreaction of the bulk fields is neglected, we consider the warp factor $\sigma(\phi)\equiv kr \abs{\phi}$ to be the RS metric where $k$ is curvature of the AdS space. We assume both quantities ($k$ and $r$) are set by the 5D Planck scale $M_\ast$ and $kr \simeq \mathcal{O}(10)$ in order to solve the gauge hierarchy problem. The flat case for an extra dimension is obtained in the limit of $k \rightarrow 0$.

As mentioned in the introduction, the aim of this work is to generate the $\mathbb{Z}_2$-odd bulk fermion masses dynamically through Yukawa interactions with a $\mathbb{Z}_{2}$-odd scalar field $\Sigma(x,\phi)$ which develops a $\phi$-dependent VEV.  The scalar action for the $\mathbb{Z}_{2}$-odd real bulk scalar on a $\mathbb{R}^{3,1}\otimes S_1/\mathbb{Z}_2$ spacetime reads
\begin{equation}\label{eq:5Daction}
S_{\rm 5D}^{\rm scalar} = \int \!d^4x \int_{-\pi}^\pi \!d\phi \sqrt{g} \; \Big\lbrace \dfrac{1}{2} g^{MN} (\partial_{M} \Sigma) (\partial_{N} \Sigma) -V(\Sigma) -\sum_i \frac{\sqrt{\abs{\hat g_i}}}{\sqrt{g}}V_i(\Sigma)\delta(\phi-\phi_i)\Big\rbrace ,
\end{equation}
where $g$ is the determinant of the 5D metric, $\hat g_i$ is the determinant of 4D metric defined on the D3-branes at $\phi_{i} = 0,\pm \pi,$ and $V_{i}$ are possible brane-localized potentials. Note that the bulk scalar field $\Sigma$ is $\mathbb{Z}_{2}$-odd w.r.t.\ the orbifold fixed points, therefore it must vanish at both brane locations. Hence, the brane potentials $V_i(\Sigma)$ could only be constants, which we choose to be the same as in the original RS setup~\cite{Randall:1999ee}. As we will see later, the bulk scalar potential $V(\Sigma)$  induces a background solution $\omega(\phi)$ for the scalar field in the bulk, i.e., a $\phi$-dependent VEV. 
Taking this into account, the scalar field can be defined as
\begin{equation}
\Sigma (x,\phi) = \ome + \dfrac{e^{\siphi}}{\sqrt{r}} \Ss ,		\label{def_bulkscalar}
\end{equation}
where $\Ss$ is the scalar fluctuation about the VEV. Now we can rewrite the above scalar action as
\begin{equation}
\begin{split}
S_{\rm 5D}^{\rm scalar} = & \; 2  \int d^4x \int_{0}^\pi d\phi \, e^{-4\sigma} \; \bigg\lbrace \dfrac{1}{2} e^{4\sigma} \; \partial_{\mu} S \, \partial^\mu S - \dfrac{1}{2r} \bigg[ \dfrac{1}{r} \Big(\partial_{\phi} \, e^{\sigma} S \Big)^2 + \; (\partial_{\phi} \, \omega)^2 \bigg] \\
	& +  \dfrac{e^{5\sigma}}{r^{3/2}} \, S\; \partial_{\phi} \Big( e^{-4\sigma} \, \partial_{\phi} \, \omega \Big) \,  - r V (\Sigma) \bigg\rbrace - \int d^4x \,  e^{-4\sigma} \bigg\lbrace \dfrac{2 e^{\sigma}}{r^{3/2}}\, S \; \partial_{\phi} \, \omega + \sum_i V_i \bigg\rbrace \bigg\vert^{\pi}_{\phi=0} .
\label{S5daction}
\end{split}
\end{equation}
In the tadpole term an integration by parts has been performed. However, since $\Sigma(x,\phi)$ is odd, the corresponding boundary terms simply vanish. We now introduce the explicit form for the scalar potential
\beq
V (\Sigma) = \lb + \dfrac{\mu^2}{2} \Sigma^2 + \dfrac{\lambda}{4!}\Sigma^4 ,
\label{mhpot}
\eeq
where $\lb$ is the bulk cosmological constant and $\mu^2<0$ is demanded for non-trivial background solutions being realized by the scalar field. From here on, we use the notation $\abs{\mu}\equiv \sqrt{\abs{\mu^2}}$. The parameters have mass dimension $[\mu] = 1$ and $[\lambda] = -1$, while the scalar field has mass dimension $[\Sigma] = 3/2$ (which implies $[S] = 1$ and $[\omega] = 3/2$).

We can now obtain the Euler-Lagrange equations for the background field $\ome$ by using the variational principle:
\beq
\omega''(\phi) - 4  \sigma'(\phi)\,  \omega'(\phi) - r^2 \frac{\partial V(\Sigma)}{\partial\omega}\Big\vert_{S=0} = 0 , 
\label{eomome}
\eeq
where primes denote derivatives w.r.t.\ the fifth-coordinate $\phi$. It can be easily seen that the trivial ansatz, $\ome = 0$, is a solution of \eqref{eomome}. However, our goal is to find out if there exists a non-trivial solution with a lower energy, while still satisfying the Dirichlet boundary conditions 
\beq
\omega(0) = \omega(\pi) = 0.
\label{bcsome}
\eeq
One can define a 4D energy density from the Hamiltonian associated with the scalar action \eqref{S5daction}. Using the equation of motion \eqref{eomome}, we find 
\beq
\rho_{E} = - \dfrac{\lambda r}{12} \int_{0}^\pi d \phi \,e^{-4 \sigma(\phi)} \omega^4(\phi) .
\label{eq:ener1}
\eeq
This implies that if a non-trivial VEV is developed, it will always correspond to a lower energy state than that of the trivial solution $\ome = 0$.

\begin{figure}[t!]
\begin{center}
\includegraphics[scale=1]{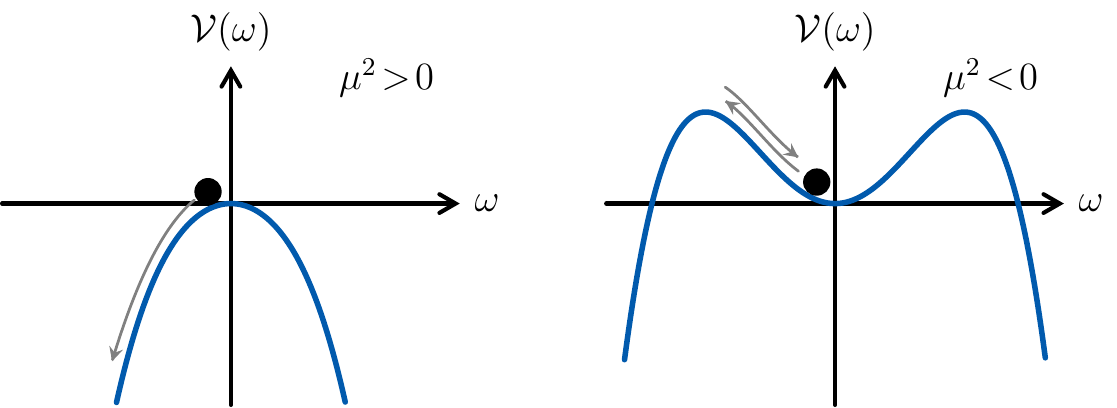}\vspace{5pt}
\caption{Illustration of the upside-down potential ${\cal V}(\omega)$ for $\mu^2  > 0$ (left) and $\mu^2 < 0$ (right).}
\label{fig:cond}
\end{center}
\end{figure}

In order to understand the dynamics leading to a non-trivial background solution for the bulk odd scalar field $\Sigma$, we need to solve \eqref{eomome}. This is the equation of a damped anharmonic oscillator, where however the damping term has the wrong sign for $\sigma'(\phi)>0$. To fix this, we redefine the coordinate of the orbifold to $\varphi\equiv \pi - \phi$, such that the UV and IR branes are located at $\varphi=\pi$ and $\varphi=0$, respectively.  Without loss of generality we restrict $\varphi\in [0,\pi]$. Denoting derivatives with respect to $\varphi$ by  dots, we can rewrite \eqref{eomome} with \eqref{mhpot} as
\beq
\ddot{\omega}(\vp) - 4 \dot\sigma(\vp)\, \dot{\omega}(\vp) + \abs{\mu r}^2 \omega(\vp) - \dfrac{\lambda r^2}{6}\omega^3(\vp) = 0 ,
\label{eomome2}
\eeq
with the damping term  having opposite sign in the RS case, $\dot{\sigma}(\vp)=-kr$, while  the boundary conditions remain unchanged. This equation of motion can be understood as
\beq
\ddot{\omega}(\vp) =  4 \dot\sigma(\vp)\, \dot{\omega}(\vp) - r^2 \dfrac{d {\cal V}(\omega)}{d \omega} ,
\label{eomome3}
\eeq
where we have defined the upside-down potential
\beq
{\cal V} (\omega) \equiv - \dfrac{\mu^2}{2} \omega^2 - \dfrac{\lambda}{4!}\omega^4 = - \, V(\Sigma)\big\vert_{S=0}\,. \label{UDpot}
\eeq
Equation~\eqref{eomome2} describes the damped motion of a particle in the potential ${\cal V}(\omega)$, see figure~\ref{fig:cond}. The boundary conditions in \eqref{bcsome} imply that the particle starts at the origin ($\omega=0$) at ``time'' $\varphi=0$ and returns to the origin after a time $\varphi=\pi$. For positive $\mu^2$ the only solution satisfying the boundary conditions is the trivial one $\omega(\varphi)=0$, since otherwise the particle will roll down the potential and never return. Having a non-trivial solution thus requires $\mu^2<0$, as might have been expected from the start. However,  as we will see later, this condition is not yet sufficient in the case of non-zero curvature.

It is instructive to rewrite \eqref{eomome2} through the rescaling
\beq
\omega(\vp) = \sqrt{\dfrac{6 \abs{\mu^2}}{\lambda}} \, v(\vp),
\label{normfact1}
\eeq
such that the quartic coupling $\lambda$ disappears, becoming just a part of normalization factor. In terms of the dimensionless field $v(\vp)$, relation \eqref{eomome2} becomes
\beq
\ddot{v}(\vp) - 4 \dot\sigma(\vp)\, \dot{v}(\vp) + \abs{\mu r}^2 \left[ v(\vp) - v^3(\vp) \right] = 0 .
\label{eomv1}
\eeq
Note that the maximum of the inverted potential is now located at $v = \pm 1$. Moreover, the 4D energy density~\eqref{eq:ener1} takes the following form
\beq
\rho_{E}  = - 3 r \dfrac{\abs{\mu}^4}{\lambda} \int_{0}^\pi d \varphi \,e^{-4 \sigma} v^4(\varphi).
\label{eq:ener2}
\eeq
In the following two subsections we consider two special cases where $\sigma(\varphi)=0$, i.e.\ the flat extra dimension, and $\sigma(\varphi)=kr(\pi -\varphi)$, the RS warp factor (AdS geometry). In section~\ref{warped_br} we employ a general warp function $\sigma(\varphi)$, which takes into account the backreaction due to the bulk scalar field. 

\subsection{Flat extra dimension}

\begin{wrapfigure}{r}{0.37\textwidth}
\centering
\includegraphics[scale=1]{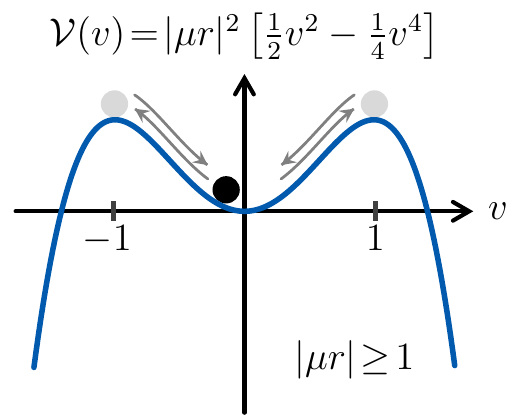}
\caption{Sketch of the potential ${\cal V}(v)$ and the motion of the particle in the flat case.}
\label{fig_potflat}
\end{wrapfigure}

In order to build up  intuition about the structure of the solutions for the VEV of an odd bulk scalar field, it is useful to review first the case of a flat metric,  $\sigma(\vp)=0$, see also~\cite{Georgi:00,Grzadkowski:2004mg,Toharia:2007xe}. In this case $\dot\sigma=0$, and therefore the damping term in \eqref{eomv1} vanishes, i.e.\
\beq
\ddot{v}(\vp) + \abs{\mu r}^2 \left[ v(\vp) - v^3(\vp) \right] = 0 .
\label{feomv1}
\eeq
This equation describes the undamped motion of a particle exposed to a non-linear conservative force. In the absence of the quartic term in the potential the motion would be harmonic, 
\beq
v(\vp) = N \sin \left( \abs{\mu r} \vp \right) ,
\eeq
where $N$ is an arbitrary normalization constant. The boundary condition at $\vp = 0$ is trivially satisfied. However, the boundary condition at $\vp = \pi$ is only satisfied for the very special parameter values $\abs{\mu r} = n \in N$ and therefore only specific values of $\mu$ are solutions of our boundary-condition problem. However, the anharmonic term in~\eqref{feomv1} has the effect of slowing down the motion as the particle climbs further up the potential, and hence the boundary condition at $\vp = \pi$ can {\it always} be satisfied as long as $\abs{\mu r} \geq 1$, see figure~\ref{fig_potflat}. More generally, if $[\abs{\mu r}] = n$ with some integer $n \geq 1$, with $[x]$ denoting the largest integer smaller than $x$, there will be exactly $n$ solutions of~\eqref{feomv1} satisfying the boundary conditions, and the one with the lowest energy density is the one with the largest amplitude, corresponding to the motion where the particle climbs exactly once up the potential and has a turning point at $\vp = \pi/2$.

In the flat case we can solve for the motion explicitly, using the energy conservation law (a first integral of the equation of motion)
\beq
E = \dfrac{1}{2} \dot{v}^2(\vp) + \dfrac{\abs{\mu r}^2}{2} \left[ v^2(\vp) - \dfrac{1}{2} v^4(\vp)\right] = \dfrac{\abs{\mu r}^2}{2} \left[ v_{m}^2 - \dfrac{1}{2} v_{m}^4\right] ,
\eeq
where $v_{m}< 1$ is the height of the turning point. Note that we cannot have $v_{m} > 1$, as in that case the particle would roll over the maximum of the inverted potential and roll away to infinity, thus violating the boundary condition at $\vp = \pi$. For $v_{m} = 1$ the particle would come to rest at the maximum of the inverted potential and stay there forever, violating as well the boundary condition at $\vp = \pi$. 

We can integrate the above first-order differential equation, obtaining
\beq
\abs{\mu r} \vp = \int_{0}^{\frac{v(\vp)}{v_{m}}} dx \frac{1}{\sqrt{1-x^2 - \dfrac{v_{m}^2}{2}(1-x^4)}} = \sqrt{\dfrac{2}{2-v_{m}^2}} \, F \left( \arcsin \dfrac{v(\vp)}{v_{m}} \right\vert \left. \dfrac{v_{m}^2}{2-v_{m}^2} \right)
\label{flatsol1}
\eeq
where $F(x \vert z)$ is the elliptic integral of the first kind. The condition that $\vp = \pi/2$ at the
   turning point, i.e.\ where $v = v_{m}$, yields
\beq
\abs{\mu r} \, \dfrac{\pi}{2} = \sqrt{\dfrac{2}{2-v_{m}^2}} \, K \left(\dfrac{v_{m}^2}{2-v_{m}^2} \right) ,
\label{flatsol2}
\eeq
where $K(z) = F(\frac{\pi}{2} \vert z)$ is the complete elliptic integral of the first kind. In the limit where $\abs{\mu r} \gg 1$, relation~\eqref{flatsol2} implies
\beq
v_{m} \approx 1-4 \exp \left( -\dfrac{\pi}{\sqrt{2}} \abs{\mu r}\right) ,
\eeq
up to exponentially suppressed terms. Solution~\eqref{flatsol1} can then be approximated as (for $0 \leq \vp < \pi/2$)
\beq
\abs{\mu r} \, \vp  \approx \atanh \dfrac{v(\vp)}{v_{m}} , 
\eeq
and the reversed motion applies for $\pi/2 < \vp \leq \pi$. Using~\eqref{normfact1}, we recover for the original VEV Georgi~{\it et~al.}'s approximate solution~\cite{Georgi:00}
\beq
\ome \approx \sqrt{\dfrac{6 \abs{\mu^2}}{\lambda}} \tanh \left( \dfrac{\abs{\mu r}}{\sqrt{2}} \phi \right) \tanh \left( \dfrac{\abs{\mu r}}{\sqrt{2}} (\pi - \phi ) \right) ,
\label{Georgieq}
\eeq
which holds up to exponentially small terms.

\subsection{Warped space}
\label{warped_sol_wobr}

We now turn to the case of a warped extra dimension and consider the RS metric with $\sigma(\vp)= kr(\pi-\vp)$. In this geometry the equation of motion for the background field $\ome$ takes the form
\beq
\ddot{\omega}(\vp) + 4 k r \, \dot{\omega}(\vp) + \abs{\mu r}^2 \omega(\vp) - \dfrac{\lambda r^2}{6}\omega^3(\vp) = 0 ,
\label{weomome2}
\eeq
where the damping term is now a linear function of $\dot{\omega}$. We can redefine $\omega(\varphi)=e^{-2\sigma}\,\bar \omega(\varphi)$, such that the single derivative term in the above equations is canceled. Therefore, the equation of motion for $\bar \omega$ reads
\begin{equation}\label{eomomebar}
   \ddot{\bar \omega} = (\mu^2+4k^2)\,r^2\,\bar \omega + \frac{\lambda r^2}{3!}\,e^{-4\sigma}\,\bar \omega^3 \,,
\end{equation}
which is analogous to the flat case of previous subsection.
Analyzing solutions of \eqref{eomomebar} with small amplitude one finds that due to the damping term oscillations are possible only if $\mu^2 + 4k^2 < 0$.  This is only a necessary, not a sufficient condition, see  (\ref{condback2}) below. For $-4k^2<\mu^2<0$ there are no oscillations but exponentially damped motion, in which the particle returns to the origin only at infinite time.   

Note  that the {\it effective} mass parameter $(\mu^2 + 4k^2)$ in~\eqref{eomomebar} is precisely the one appearing in the Breitenlohner-Freedman (BF) bound in AdS space~\cite{BF:82-1,BF:82-2}. The BF bound $\mu^2 + 4k^2\geq0$ is related to the stability of the AdS space, i.e.\ the absence of tachyonic modes of the scalar fluctuations. Moreover, the AdS/CFT duality~\cite{Maldacena:1997re, Gubser:1998bc} implies that a theory with a bulk scalar field in AdS space has a dual strongly coupled conformal field theory (CFT). In particular, in the usual RS setup with two branes the AdS/CFT duality implies that the dual CFT has a UV cutoff at $\Lambda_{\textsc{uv}}$ due to the presence of the UV brane, and that 4D gravity is dynamical~\cite{ArkaniHamed:2000ds}. This CFT remains unbroken from $\Lambda_{\textsc{uv}}$ down to an energy scale $\Lambda_{\textsc{ir}}$, where it is spontaneously broken by the presence of the IR brane in the 5D theory~\cite{Rattazzi:2000hs}. It was shown in~\cite{Klebanov:1999tb} that the saturation of the BF bound implies the merging of two 4D CFTs with operators of scaling dimensions $\Delta_+$ and $\Delta_-$ in the dual field theory, where $\Delta_\pm=2\pm \sqrt{4 - |\mu|^2/k^2}$. Furthermore, it was shown in~\cite{Pomoni:2008de} that the violation of the BF bound leads to imaginary scaling dimension for the CFT operators, which signals the breakdown of the dual CFT. Note, however, that the BF bound is obtained for a free bulk scalar theory, whereas in our case the scalar field has quartic interactions leading to the appearance of a non-trivial VEV in the bulk. As we argued above, the violation of the BF bound is a necessary condition for this to happen. Only in this case, the bulk scalar field condenses and acquires a $\phi$-dependent VEV, which leads to emergence of a scale related to the bulk scalar and the loss of conformality of the dual theory~\cite{Kaplan:2009kr}. We leave a more detailed discussion of the dual theory for our setup with a $\mathbb{Z}_2$-odd bulk scalar field for future work.

It is instructive to recast~\eqref{weomome2} in terms of $v(\vp)$, such that in the RS geometry we get
\beq
\ddot{v}(\vp) + 4 k r \, \dot{v}(\vp) + \abs{\mu r}^2 \left[ v(\vp) - v^3(\vp) \right] = 0 .
\label{weomv1}
\eeq
In the absence of the quartic term in the potential the motion would be a damped harmonic oscillation,
\beq
v(\vp) = N e^{-2 k r \vp} \sin (\nu\, \vp ) , 
\eeq
where the frequency $\nu \equiv \sqrt{\abs{\mu r}^2 - 4(kr)^2}$ is real and $N$ is again an arbitrary normalization factor.  While the boundary condition at $\vp = 0$ is trivially satisfied, the boundary condition at $\vp = \pi$ is only satisfied for the very special parameter values $\nu = n \in N$. By the same arguments as for the flat case, now for another definition of $\nu$, the boundary condition at $\vp = \pi$ can always be satisfied as long as $\nu \geq 1$. More generally, for $[\nu]=n$ with some integer $n\ge 1$ there will be exactly $n$ solutions of~\eqref{weomv1} satisfying the boundary conditions, and the one with the lowest energy density~\eqref{eq:ener2} is the one with the largest amplitude, corresponding to the motion where the particle climbs exactly once up the potential and falls down, satisfying the boundary condition at $\vp = \pi$. Note that in this case the turning point will not be placed at $\vp = \pi/2$ but will be slightly shifted onto a lower value of $\vp$ due to the damping.  It follows from this discussion that the correct condition for having a non-trivial background solution reads  
\beq
\nu^2=\abs{\mu r}^2 - 4(kr)^2 > 1 .
\label{condback2}
\eeq
\begin{figure}[t!]
\begin{center} \hspace*{-0.65cm}
\includegraphics[scale=0.42]{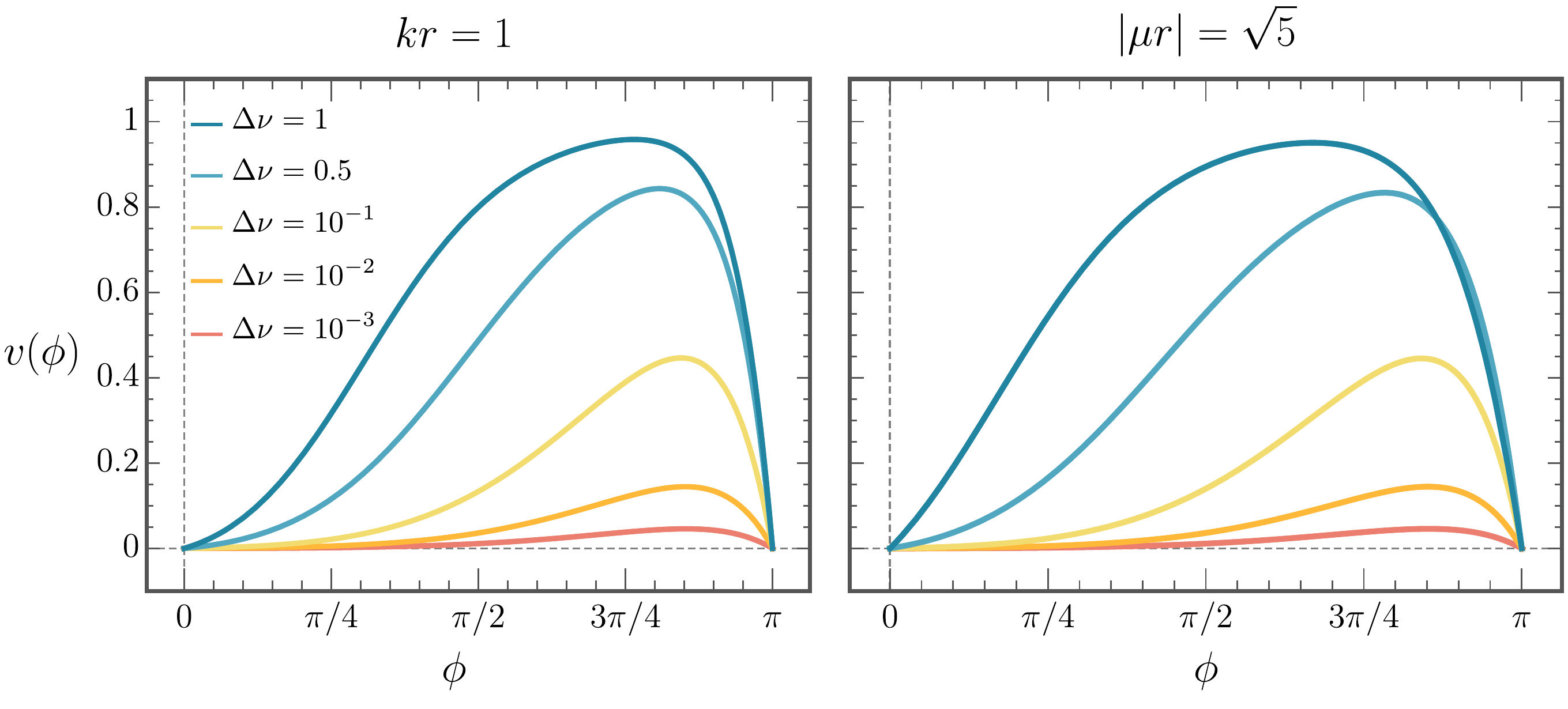} 
\caption{Solutions for the scalar VEV $v(\phi)$, with $\phi=\pi-\varphi$, for different choices of $\Delta\nu=\nu-1$ and fixed values of $kr$ (left) and $\abs{\mu r}$ (right).}
\label{fig:vevsol1}
\end{center}
\end{figure}

\begin{figure}[t]
\begin{center} \hspace*{-0.65cm}
\includegraphics[scale=0.42]{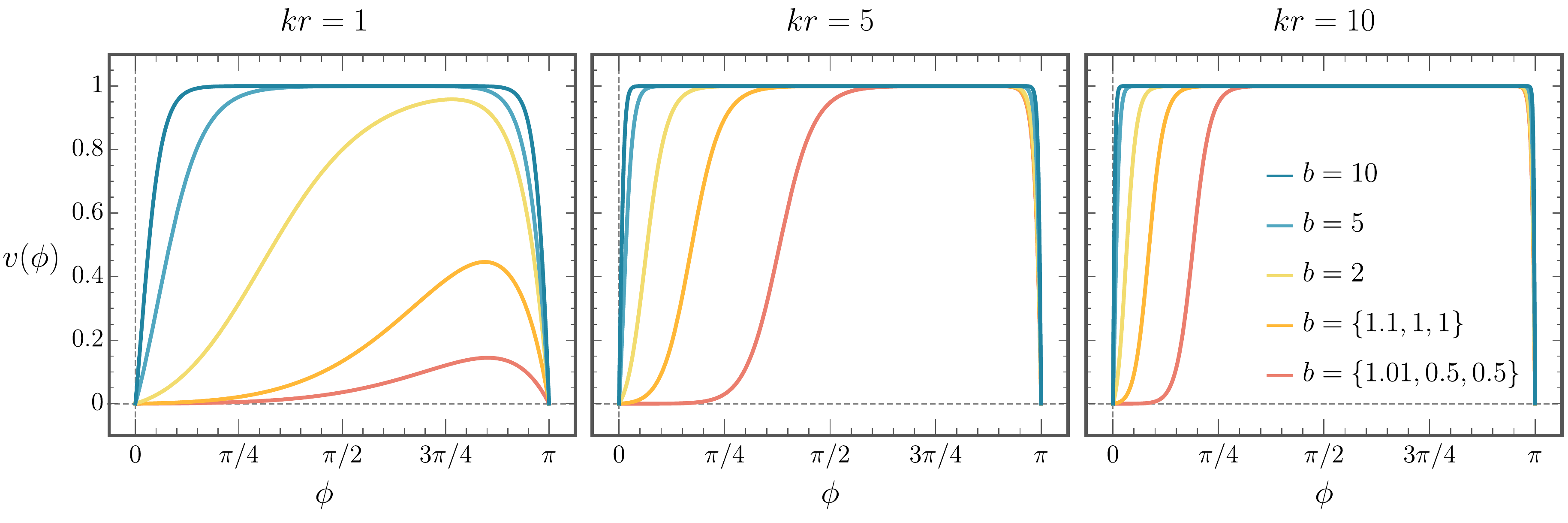} 
\caption{Solutions for the scalar VEV $v(\phi)$ for different choices of $kr$ and $b>1/kr$. In the legend, we show in brackets the values of $b$  used for the different figures, from left to right, such that the constraint $b>1/kr$ is satisfied.}
\label{fig:vevsol2}
\end{center}
\end{figure}

In a natural setup one would assume that the dimensionless quantities $\abs{\mu r}$ and $kr$ are chosen to be of $\mathcal{O}(1)$, leading therefore to $\nu=\mathcal{O}(1)$. Some representative solutions for the scalar VEV $v(\phi)$ in this case are shown in figure~\ref{fig:vevsol1} for fixed values of $kr$ (left) and $\abs{\mu r}$ (right) and different values of $\Delta\nu\equiv\nu-1$. Note that the maximum possible value of the VEV, $v_{m}=1$, is only approached for values of $\Delta\nu\gtrsim 1$, whereas the amplitude of the oscillation gets smaller as $\Delta\nu$ gets closer to zero. One can also see that small values of $\Delta\nu$ not only result in smaller amplitudes, but also lead to very asymmetric VEV profiles, which look rather different from a step function. In realistic versions of RS models, the product $kr$ is chosen such that the hierarchy problem is solved, leading to the condition that $L\equiv kr\pi\approx 33$. In this case the quantities $\abs{\mu r}$ and $kr$ on the left-hand side of the bound \eqref{condback2} are both much larger than~1. It is then useful to rewrite $\nu^2\equiv (kr)^2\,b^2$, such that the bound is recast in the form
\begin{equation}\label{newbound}
   b = \sqrt{\frac{|\mu^2|}{k^2} - 4} > \frac{1}{kr} \,, 
\end{equation}
where the right-hand side is close to 0.1 for $L\approx 33$. This bound is reminiscent of the BF bound $\beta=\sqrt{\frac{\mu^2}{k^2}+4}>0$ for a free bulk scalar without spontaneous symmetry breaking. Likewise, we can rewrite the differential equation \eqref{weomv1} as
\begin{equation}
   \frac{d^2 v(y)}{dy^2} - 4\,\frac{dv(y)}{dy} + \left( 4 + b^2 \right) \left[ v(y) - v^3(y) \right] = 0 \,,
\end{equation}
where $y=kr\phi=kr(\pi-\varphi)$. It is natural to assume that the parameter $b^2$ in this equation, which is given in terms of a ratio of two Planck-scale parameters $\mu$ and $k$, is of $\mathcal{O}(1)$. The equation must then be solved with the boundary conditions $v(y_i)=0$ for $y_{\rm UV}=0$ and $y_{\rm IR}=L$. Note that the large parameter $L\approx 33$ only enters via the size of the interval on which one solves the differential equation, while the parameter $b^2$ is still naturally of $\mathcal{O}(1)$. Under these conditions, the quantity $\nu^2=(kr)^2\,b^2\gg 1$ is automatically much larger than~1, and consequently the VEV develops a profile with a broad plateau near $v_m=1$, resulting in a shape that closely resembles a step function. In figure~\ref{fig:vevsol2}, we show solutions for the scalar VEV $v(\phi)$ for different values of $b$ consistent with the bound \eqref{newbound} for $kr=1$, 5 and 10. 
While the first plot is just a different representation of figure~\ref{fig:vevsol1}, in the last two plots the presence of the large parameter $L=kr\pi\gg 1$ gives rise to a prominent plateau, whereas the behavior close to the boundaries is only controlled by the parameter $b$. 
If we plotted these results as functions of $y=k r\phi$, the behavior of the various lines in the last two plots near the two branes would be identical, and only the width of the plateau would be different.\footnote{In the mechanical analogue of figure~\ref{fig:cond}, solutions with a wide plateau correspond to fine-tuned motions where the point-mass comes to rest infinitesimally close to the maximum of the potential and stays there for a long time. The motions from the origin to the maximum and back are independent of how long the particle stays at the top.} The solutions obtained for large $kr$ indeed look much closer to the traditional case of a step function, showing that solving the hierarchy problem helps to obtain kink-like solutions for the VEV. While this is particularly true near the IR brane  
(for $\phi\lesssim\pi$), to obtain a kink-like solution also near the UV brane  
(for $\phi\gtrsim 0$) requires large values of $b$. Indeed, there is a decoupling limit, in which $b\to\infty$ and the scalar VEV $v(\phi)$ approaches $\sgn(\phi)$, such that one recovers the $\mathbb{Z}_2$-odd bulk masses of the conventional RS model.

\subsection{Warped space with backreaction on the metric}
\label{warped_br}

The presence of a non-zero VEV of the bulk scalar field has an impact on the background geometry of the warped space. In this section, we calculate this backreaction by solving the coupled scalar-gravity system, the action of which reads
\beq
S=\int d^4x \int_{-\pi}^\pi d\phi\sqrt{g}\bigg\{-\frac{\mathcal{R}}{2\kappa^2}+\frac{1}{2}g^{MN}(\partial_{M}\Sigma)(\partial_{N}\Sigma)-V(\Sigma)-\sum_i \frac{\sqrt{|\hat g_i| }}{\sqrt{g}}V_i(\Sigma)\delta(\phi-\phi_i)\bigg\},     \label{action}
\eeq
where $\mathcal{R}$ is the 5D Ricci scalar, $\kappa^{-2}\equiv2\ms^3$ with $\ms$ being the 5D Planck mass, $V(\Sigma)$ is the bulk scalar potential for scalar field $\Sigma(x,\phi)$, and $V_i(\Sigma)$ are the brane potentials.
We consider \eqref{metric} as an ansatz for the metric where $\siphi$ is a generic $\phi$-dependent warp function. Note that the 5D Planck mass $M_{\ast}$ is related to the (reduced) 4D Planck mass $M_{\rm Pl} \simeq 2\cdot 10^{15}~$TeV by 
\begin{align}
	M_{\rm Pl}^2=M_{\ast}^3 \, r \int_{0}^{\pi}d\phi\, e^{-2\sigma(\phi)}=\frac{M_{\ast}^3}{2k}\left(1-e^{-2k r \pi} \right),
\end{align}
where the second equality only holds for the RS case $\sigma(\phi)=kr|\phi|$.

Here we focus on the background solutions for the geometry and bulk scalar field, i.e.\ $\Sigma(x,\phi)=\ome$. The Einstein's equation and the equation of motion for $\ome$, resulting from the action~\eqref{action}, are 
\begin{align}
\mathcal{R}_{MN}-\frac{1}{2}g_{MN}\mathcal{R}&=\kappa^2\,T_{MN},\label{eineq1}\\
-\frac{1}{\sqrt{g}}\partial_{M}\left(\sqrt{g}g^{MN}\partial_N \omega\right)
&=\frac{\partial V(\omega)}{\partial\omega}+\sum_i \frac{\sqrt{|\hat g_i| }}{\sqrt{g}}\frac{\partial V_i(\omega)}{\partial \omega} \delta(\phi-\phi_i), \label{eineq2}
\end{align}
where the energy-momentum tensor $T_{MN}$ for $\ome$ is,
\begin{align}
T_{MN}=\partial_{M}\omega\partial_{N}\omega-g_{MN}\bigg[\frac{1}{2}g^{PQ}\partial_P\omega\partial_Q\omega-V(\omega)\bigg]+\sum_i \frac{\sqrt{|\hat g_i |}}{\sqrt{g}}V_i(\omega) \hat{g}_{\mu\nu}^{i}\delta_{M}^{\mu}\delta_N^\nu\delta(\phi-\phi_i), \label{emt}
\end{align}
with $\hat g_{\mu\nu}^i$ and $\hat g_i$ being the 4D induced metric and its determinant at the brane $i$, respectively.
From Eqs.~\eqref{eineq1} and~\eqref{eineq2}, we get the equations of motion for the warp function $\sigma(\phi)$ and scalar VEV $\omega(\phi)$ as 
\begin{align}
\sigma^{\prime2}&=\frac{\kappa^2}{6}\Big[\frac{1}{2}\omega^{\prime2}-r^2V(\omega)\Big],\label{eom01}\\
\sigma^{\prime\prime} &=\frac{\kappa^2}{3}\Big[\omega^{\prime2}+r\sum_i V_i(\omega)\, \delta(\phi-\phi_i)\Big],	\label{eom02}\\
\omega^{\prime\prime}-4\sigma^{\prime}\omega^{\prime}&=r^2\frac{\partial V(\omega)}{\partial\omega}+r\sum_i\frac{\partial V_i(\omega)}{\partial\omega}\delta(\phi-\phi_i), \label{eom03}
\end{align}
where the scalar potential has been given in~\eqref{mhpot} and includes the 5D bulk cosmological constant $\lb$. The brane localized potentials are constant brane tensions, since the odd bulk scalar field vanishes on the branes, and therefore
\beq
V_{\rm UV}(\omega)=V_{\rm UV},	\lsp V_{\rm IR}(\omega)=V_{\rm IR}.	\label{eq:potential_bt}
\eeq
These brane potentials are crucial in order to ensure that the 4D cosmological constant vanishes. The above equations have singularities due to the presence of 3-branes at $\phi=0$ and $\phi=\pi$, therefore the warped function $\siphi$ and the scalar field $\ome$ satisfy the jump conditions
\begin{align}
	\big[\sigma^\p(\phi)\big]_i=\frac{\kappa^2}{3} r V_i,	\Lsp	\big[\omega^\p(\phi)\big]_i=0,		\label{jump_cond}
\end{align}
where the  jump of a function $f(x)$ is defined as
\beq
\big[f(x)\big]\equiv \lim_{\epsilon \to 0}\Big[f(x+\epsilon)-f(x-\epsilon)\Big].
\eeq

The evolution of the fields in the bulk is described by~\eqref{eom01}--\eqref{eom03} after dropping the brane-localized terms. However, the three resulting equations are not independent.  Indeed, the background solution is completely determined by
\begin{align}
\sigma^{\prime\prime} - \frac{\kappa^2}{3} \omega^{\prime2} &= 0,	\label{eom04}\\
\omega^{\prime\prime}-4\sigma^{\prime}\omega^{\prime}-r^2\frac{\partial V(\omega)}{\partial\omega}&=0. \label{eom05}
\end{align}
Note that $\sigma(\phi)$ only appears through its derivatives both in~\eqref{jump_cond} and~\eqref{eom04}--\eqref{eom05}. Hence, we have a system of two coupled differential equations of first order in $\sigma'(\phi)$ and second order in $\omega(\phi)$, which can be  solved with three integration constants. However, note that we have four constraints coming from boundary conditions and the jump conditions~\eqref{jump_cond}. Three of these constraints fix the three integration constants, whereas the remaining one corresponds to the usual fine-tuning required to make the ansatz~\eqref{metric} a solution, which guarantees in particular a vanishing 4D cosmological constant. 

The relations between the 5D bulk cosmological constant and brane tensions, which is required by 4D Poincaré invariance, gets modified in our case  compared with the original RS model.  In particular, for a non-trivial scalar VEV the RS relation  $V_{\rm UV}^{\rm RS}=-V_{\rm IR}^{\rm RS}=6k / \kappa^2 $ is in conflict with the bulk equation of motion~\eqref{eom04}. This can be seen by integrating \eqref{eom04} over the extra dimension $\phi$. One obtains (with $x_\pm\equiv x\pm\epsilon$)
\beq
\sigma' (\pi_-) = \sigma' (0_+) +  \frac{\kappa^2}{3} \int_{0_+}^{\pi_-}  \!d\phi \, \omega'^2 (\phi) ,  \label{nonRScond}
\eeq
which implies that the RS solution with $\sigma' (\pi_-) = \sigma' (0_+) = kr$ will no longer solve the system of equations in the presence of a non-trivial VEV. As a consequence,  the first jump condition in~\eqref{jump_cond} implies that the two brane tensions
\beq
V_{\text{UV}}=\frac{6 \sigma^\p(0_+)}{r\kappa ^2 },		\lsp V_{\text{IR}}=-\frac{6 \sigma^\p(\pi_-)}{r\kappa ^2}
\eeq
are not longer equal and opposite as in the RS case. One can fix one of the integration constants by choosing the boundary value of the warp function on the UV brane, such that in the UV the background geometry asymptotes to the AdS space, i.e.\ $\sigma^{\prime}(0_{+})=kr$. Hence, the brane tensions in our model are
\beq
V_{\text{UV}}=\frac{6 k}{\kappa ^2 },		\lsp V_{\text{IR}}=-V_{\text{UV}}-\frac{2}{r} \int_{0_+}^{\pi_-}  \!d\phi \, \omega'^2 (\phi) . 
\eeq
The bulk cosmological constant $\lb$ introduced in~\eqref{mhpot} can now be obtained by inserting the solutions from~\eqref{eom04}--\eqref{eom05} into~\eqref{eom01} and evaluating the fields near the UV brane, i.e.\
\beq
\sigma^{\prime2}(0_{+})=\frac{\kappa^2}{6}\Big[\frac{1}{2}\omega^{\prime2}(0_{+})-r^2 \lb\Big].
\eeq
With $\sigma^{\prime}(0_{+})=kr$ we get 
\beq
\lb = -\dfrac{6k^2}{\kappa^2} - \dfrac{\omega^{\prime2}(0_{+})}{2r^2} .
\eeq
This new contribution to $\lb$ leads to modifications of the background geometry in the IR w.r.t.\ the case of a RS space. 

Summarizing, for a non-trivial scalar VEV with Dirichlet boundary conditions we need to solve the coupled equations~\eqref{eom04} and~\eqref{eom05}, which in terms of the dimensionless VEV $v(\phi)$ take the form 
\begin{align}
	\sigma'' (\phi) - \gamma \, v'^2(\phi)&=0,		\label{eomMVEV}\\
v''(\phi) - 4 \, \sigma'(\phi) \, v'(\phi) + \abs{\mu r}^2 \left[ v(\phi) - v^3(\phi) \right] & = 0 , \label{eomMVEV2}
\end{align}
along with boundary conditions,
\beq
	v(0)=v(\pi)=0,\lsp \sigma^\p(0)=kr.			\label{eq:bc}
\eeq
Above we have defined the parameter
\begin{equation}
	\gamma \equiv \frac{\abs{\mu}^2}{\lambda M_{*}^3}=\frac{2\abs{\mu}^2\kappa^2}{\lambda},
\end{equation}
which parametrizes the scalar-gravity interaction, i.e.\ the strength of the backreaction. The limit $\gamma\to 0$ corresponds to the case of zero backreaction on the metric, where we recover the solutions of section~\ref{warped_sol_wobr}. 

The value of $\gamma$ is a free parameter of our model, which depends on a ratio involving three parameters that are naturally set by the Planck scale. On the other hand, if $|\mu|$ is significantly smaller than $M_{\ast}$ or $\lambda$ is significantly larger than $1/M_{\ast}$, values of $\gamma$ moderately smaller than 1 can be arranged without much fine tuning.  In our analysis below, we will somewhat arbitrarily consider the two cases $\gamma=0.5$ and $\gamma=0$ to represent scenarios of strong and negligible backreaction, respectively. The numerical results obtained with $\gamma\lesssim 0.1$ are very close to those found for $\gamma=0$.

\begin{figure}[t!]
\begin{center}
\hspace*{-1cm}
\includegraphics[scale = 1]{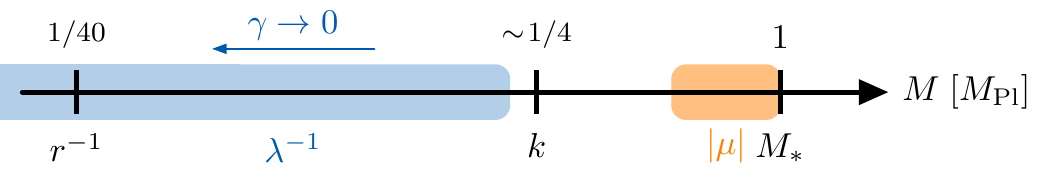}
	\caption{Illustration of the parametric values of the dimensionful parameters of our model in units of the fundamental scale $M_\ast$. Note that the small hierarchy between $k$ and $1/r\approx 0.1\, k$ is required for the model to address the gauge hierarchy problem. The relation  between the parameters $k$, $|\mu|$ and $M_{\ast}$ is in principle arbitrary, however our dynamical mechanism requires that $|\mu|>2 k$ with good approximation, see~\eqref{condback2}.} \label{fig:scales}
\end{center}
\end{figure}

Before discussing the numerical solutions of the above coupled scalar-gravity system a few comments are in order. In this work we obtain the numerical solutions by the shooting method of initial conditions. In particular, we choose the initial conditions by varying the values of $\sigma^\p(0)$ and $v^\p(0)$ at the UV brane ($\phi=0$) such that our boundary-value problem is solved. The choices of values for other parameters employed in this paper are natural in a sense that all the dimensionless parameters have $\mathcal{O}(1)$ values, whereas, the dimensionful  parameters (e.g.\ $\mu, k, \ldots$) have ${\cal O}(1)$ values in units of the 5D Planck mass $M_{\ast}$, which we may identify with the physical Planck scale $M_{\rm Pl}$. Furthermore, the value of the compactification radius $r$ of the extra dimension is chosen such that it solves the gauge hierarchy problem. We take $r = 40 M_{\text{Pl}}^{-1}$, such that $k\sim{\cal O}(33)/(\pi r)$ and $\abs{\mu} \in [20,40]/r$  are both below $M_{\rm Pl}$. The different values chosen are illustrated in figure~\ref{fig:scales}.

\begin{figure}[t!]
\begin{center}
\hspace*{-1cm}
\includegraphics[scale=0.5]{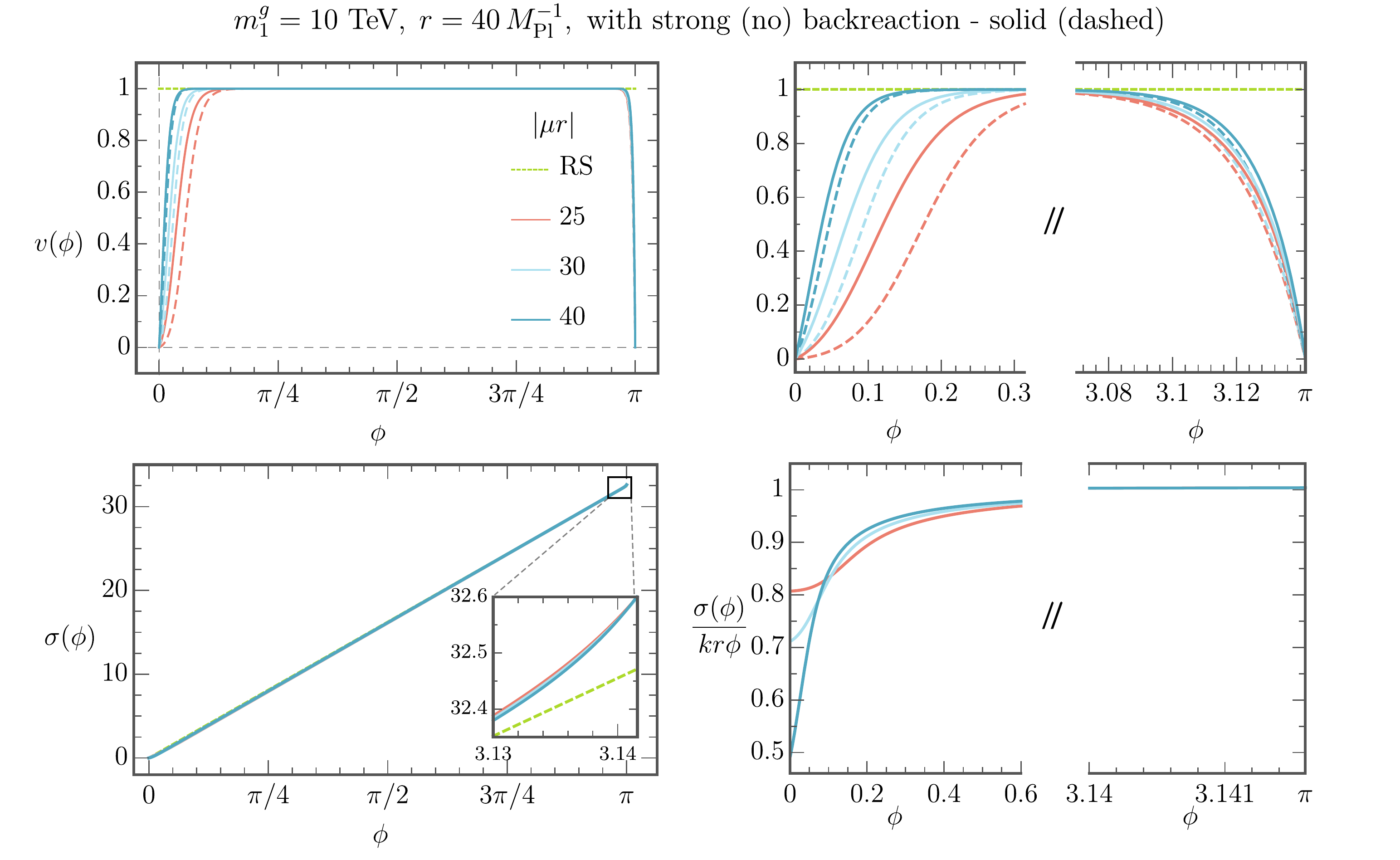} 
\caption{Background solutions for $\gamma = 0$ (no backreaction limit) and $\gamma = 0.5$ (strong backreaction) and different values of $\abs{\mu r}$. Upper panels show the scalar VEV solutions for  RS (with $\sgn(\phi)$ function) and $\abs{\mu r} = 25, 30$ and 40, with the upper right figure emphasizing the behavior of the VEV near the branes. Analogous plots are shown for the warp function $\sigma(\phi)$ in the lower panels.} \label{fig:background_sols_BR}
\end{center}
\end{figure}

In figure~\ref{fig:background_sols_BR} we show the numerical solutions for the coupled system~\eqref{eomMVEV}--\eqref{eomMVEV2}  for the case  $\gamma=0$ (no backreaction limit, dashed lines)  and $\gamma = 0.5$ (strong backreaction, solid lines), for input values of $\abs{\mu r} = 25$ (red), $30$ (light blue) and $40$ (dark blue). The upper panels show the profile of the scalar VEV $v(\phi)$ for these cases, together with the  RS case of a $\sgn(\phi)$ function (green dashed), both in the whole extra dimension (left) and in  a ``zoomed'' region near the branes (right). The lower panels show the warp function, $\sigma(\phi)$ (left), and the ratio of the warp function with respect to the RS solution, $\sigma(\phi)/kr\phi$ (right), again the latter only showing values of $\phi$ close to the branes. Note that again for $\mu\to\infty$ (corresponding to $b\to\infty$ if no tuning is assumed) we reach the decoupling limit where $v(\phi)\to\sgn(\phi)$, i.e.\ the dynamics of the bulk scalar freezes. Therefore, for a generic case with a fixed value of radius of extra dimension $r$, with increasing $\abs{\mu r}$ the solutions in the bulk get closer to the original RS case. Furthermore, it is instructive to set the scale of the model parameters in terms of some TeV-scale observable. It is common in the literature to define the so-called KK mass scale  
\beq
\mkk \equiv \dfrac{\sigma'(\pi)}{r}  e^{-\sigma(\pi)},
\label{defMKK}
\eeq
which in the context of RS models (where $\mkk = k e^{-kr\pi}$) sets the mass scale for low-lying KK excitations of the SM particles. However, as we will show in section~\ref{sec:EWPOs_NS}, once the backreaction is taken into account the ratio between e.g.\ the first vector resonance in the spectrum and $\mkk$ changes dramatically, which makes the use of $\mkk$ for phenomenological studies problematic.  Therefore, in this work we will use the mass of the first KK gluon, $m_1^g$, to define the scale of new physics. For all solutions shown in figure~\ref{fig:background_sols_BR} this mass is set to $m_1^g = 10$ TeV.

\subsubsection*{Comments on the stability of the warped extra dimension}
\label{stability_comments}

We briefly comment on the issue of stability of the extra dimension with an odd bulk scalar. In any realistic warped extra-dimensional model which solves the gauge hierarchy problem it is essential to have a dynamical mechanism to stabilize the radius $r$ of the extra dimension, i.e.\ the distance between the two branes. Without such a mechanism, the quantum perturbation of the $55$ component of the metric~\eqref{metric}, the so-called {\it radion}, remains massless. However, as shown by Goldberger and Wise (GW)~\cite{Goldberger:1999uk}, the radius of extra dimension can be dynamically stabilized by a 5D bulk scalar which has non-trivial potentials in the bulk and at both branes. Given the fact that the bulk scalar field  $\Sigma(x,\phi)$, which we have introduced in our model  to generate the fermion bulk masses, affects the background geometry in a non-trivial way, it is natural to ask whether such a bulk scalar could also serve as a stabilizing GW field. 

In order to answer this question, one needs to solve Einstein equations to linear order in perturbations of the metric as well as the bulk scalar field, as e.g.\ in~\cite{Csaki:2000zn}. After this exercise, one ends up with a bulk scalar, which is a linear combination of metric and scalar field perturbations, and it has a zero mode, corresponding to the light radion, and a tower of KK modes. However, it was shown in~\cite{Lesgourgues:2003mi} that a scalar field with an extremum point in the bulk of the extra dimension cannot stabilize the radius, in fact, it develops a tachyonic mode which introduces an instability in the system. We have explicitly verified this result and found a tachyonic mode in the spectrum of scalar perturbations. Hence, an odd bulk scalar alone cannot stabilize the extra dimension. However, one can argue that introducing a second bulk scalar as a GW stabilizing field can not only stabilize the size of extra dimension but also remove the tachyonic mode from the scalar spectrum. This analysis is beyond the scope of the present work, but we note that partial progress has been made along these lines in~\cite{Toharia:2010ex}. Hence, in what follows we consider the backreaction of the background solutions, however, assume the geometry to be static (for both weak and strong backreaction), i.e.\ we do not consider the scalar perturbations of the metric but only consider the fluctuations of the odd bulk scalar around its background VEV as in \eqref{def_bulkscalar}, i.e.\ $\Sigma (x,\phi) = \ome + e^{\siphi}\Ss/\sqrt{r}$.


\section{Gauge bosons and electroweak precision tests}
\label{sec:gauge}

In this section we calculate the impact on tree-level electroweak precision observables due to the backreaction from the $\mathbb{Z}_2$-odd bulk scalar on the metric. We consider both the minimal RS case and its custodial extension in the presence of a brane-localized Higgs sector. 

\subsection{Minimal case}
We closely follow the treatment of the KK decomposition for the electroweak sector from~\cite{Casagrande:2008hr, Casagrande:2010si}.  We start by considering a $SU(2)_L\times U(1)_Y$ bulk gauge group, spontaneously broken to $U(1)_Q$ by the VEV of an IR-localized Higgs. In order to have SM gauge bosons as the zero modes of bulk gauge bosons, we require the vector components $W_\mu^a$ and $B_\mu$ to be even and the scalar components $W_\phi^a$ and $B_\phi$ to be odd under the $\mathbb{Z}_2$ orbifold symmetry, respectively. We write the gauge-sector bulk action as
\beq
S^{\rm gauge}_{\rm 5D} = \int d^4x\int_{-\pi}^\pi\!d\phi\, \sqrt{g}\,g^{KM} g^{LN}\Big(\! - \frac14\,W_{KL}^a W_{MN}^a - \frac14\,B_{KL} B_{MN}  \Big)+S_{\rm Higgs} + S_{\rm GF} + S_{\rm FP} \,,	\label{eq:gauge_action_1}
\eeq
where $W^a_{MN}$ and $B_{MN}$ are the 5D field strength tensors for $SU(2)_L$ and $U(1)_Y$. $S_{\rm GF}$ and $S_{\rm FP}$ are the gauge fixing and the Faddeev-Popov ghost actions, respectively. For simplicity, the Higgs-sector is localized on the IR brane. The action for the canonically normalized Higgs doublet $H(x)$ reads
\begin{equation}
	S_{\rm Higgs} = \int d^4x \, \Big[ (D_\mu H)^\dagger\,(D^{\mu} H) + (e^{-\sigma(\pi)}\mu_H)^2 H^\dagger H + \lambda_H \left(  H^\dagger H \right)^2 \Big] ,
\end{equation}
where, after the electroweak symmetry breaking, $H(x)$ can be decomposed as
\beq    
H(x) = \frac{1}{\sqrt2}
   \left( \begin{array}{c}
    -i\sqrt2\,G^+(x) \\
    v_{h} + h(x) + iG^3(x)
   \end{array} \right) ,
\eeq
with $v_{h}\simeq 246$\,GeV the SM Higgs VEV, $h(x)$ is the Higgs field and $G^\pm$ and $G^3$  the associated Goldstone bosons. The covariant derivative $D_\mu$ is defined as
\beq
D_\mu =\partial_\mu-i\frac{g_5}{2} \tau^a W^a_\mu -i\frac{g_5^\p}{2} B_\mu,      \label{co_dir_M_SM}
\eeq
with $\tau^a$ the Pauli matrices, and $g_5$ and $g^\p_5$  the 5D gauge couplings of $SU(2)_L$ and $U(1)_Y$, respectively.

We write the KK decomposition of the 5D gauge fields in the form of 4D vector bosons and scalars as~\cite{Casagrande:2008hr}
\begin{equation}
\mathbb{A}_\mu(x,\phi) = \frac{1}{\sqrt r} \sum_n \mathbb{A}_\mu^{(n)}(x)\,\chi_n^\mathbb{A}(\phi) \,, \qquad 
\mathbb{A}_\phi(x,\phi)= \frac{-1}{\sqrt r} \sum_n \frac{1}{m_n^\mathbb{A}}\,\mathbb{A}_\phi^{(n)}(x)\, \partial_\phi\,\chi_n^\mathbb{A}(\phi) \,, \label{eq:gauge_decomp}
\end{equation}
where $\mathbb{A}=A,Z,W^{\pm}$ are the usual field redefinitions of the electroweak gauge bosons 
\begin{align}
   W_M^\pm &= \frac{1}{\sqrt2} \left( W_M^1\mp i W_M^2 \right) , 
   \\
   Z_M &= \frac{1}{\sqrt{g_5^2+g_5^{\prime 2}}} 
    \left( g_5 W_M^3 - g_5^\p  B_M \right) , \\
   A_M &= \frac{1}{\sqrt{g_5^2+g_5^{\prime 2}}} 
    \left( g_5^\p  W_M^3 + g_5 B_M \right) . 
\end{align}
Above $\mathbb{A}_{\mu/\phi}^{(n)}(x)$ are the 4D KK mass eigenstates, and the dimensionless $\chi_n^\mathbb{A}(\phi)$ are the KK wave-functions, which form complete sets on the orbifold. These KK wave-functions satisfy the orthonormality condition
\beq
   2\int_{0}^\pi\!d\phi\,\chi_m^{\mathbb{A}}(\phi)\,\chi_n^{\mathbb{A}}(\phi)
   = \delta_{mn} \,. \label{chinorm}
\eeq
Note that both the kinetic terms for the gauge fields and the Higgs field in the action~\eqref{eq:gauge_action_1} contain mixed terms involving the gauge bosons with their scalar companions and the Goldstone fields, respectively. These mixings can be removed by introducing appropriate gauge-fixing terms in the Lagrangian, see for instance~\cite{Casagrande:2008hr}. Inserting the KK decompositions into the action~\eqref{eq:gauge_action_1} leads to the equation of motion 
\beq
   - \frac{1}{r^2}\,\partial_\phi\Big(e^{-2\sigma(\phi)}\,
   \partial_\phi\,\chi_n^\mathbb{A}(\phi) \Big)= (m_n^\mathbb{A})^2\,\chi_n^\mathbb{A}(\phi) 
   \,,\label{eq:gauge_eom}
\eeq
along with the boundary conditions 
\beq
\partial_\phi\,\chi_n^{\mathbb{A}}(0) = 0 \,, \lsp \Big[\partial_\phi + \pi r^2\tilde{m}_{\mathbb{A}}^2  e^{2\sigma(\pi)} \Big] \chi_n^{\mathbb{A}}(\pi^-) = 0. \label{eq:gauge_bcs}
\eeq
Above $\tilde{m}_{\mathbb{A}}$ corresponds to the mass arising from the interaction with the brane-localized Higgs field, with
\begin{align}
	\tilde{m}_W^2=\frac{g_5^2}{2\pi r} \frac{v_{h}^2}{4 }, \qquad \tilde{m}_Z^2=\frac{(g_5^2+g_5^{\prime 2})}{ 2\pi r}\frac{v_{h}^2}{4 }, \qquad \tilde{m}_A^2=0.
\end{align}
The equations for the gluon case are the same as for the photon. For $\mathbb{A}=Z,W^{\pm},$ we obtain the following expressions for the physical zero-mode masses 
\begin{align}
	m_0^{\mathbb{A},2}&= \tilde{m}_{\mathbb{A}}^2\left[1-\frac{r^2\tilde{m}_{\mathbb{A}}^2}{\pi}\int_0^\pi d\phi_1 e^{2\sigma(\phi_1)}\phi_1^2+\mathcal{O}\left(\frac{v_{h}^4}{M_{\rm KK}^4}\right)\right],
	\label{eq:mass0}
\end{align}
while the photon remains massless as expected. 

The  momentum-space 5D gauge boson propagator for $\mathbb{A}_{\mu}$ is defined by 
\beq
   \left[\frac{1}{r^2}\,\partial_\phi e^{-2\sigma(\phi)}\,\partial_\phi +\hat{p}^2 \right]D_\mathbb{A}(\phi,\phi^{\prime};\hat{p})=\delta(\phi-\phi^{\prime}),
\eeq
along with the boundary conditions
\beq
\partial_{\phi}D_{\mathbb{A}}(\phi,\phi^{\prime};\hat{p})\Big\vert_{\phi=0}=0, \lsp  \Big[\partial_\phi + \pi r^2\tilde{m}_{\mathbb{A}}^2  e^{2\sigma(\pi)} \Big] D_{\mathbb{A}}(\phi,\phi^{\prime};\hat{p})\Big\vert_{\phi=\pi^-}=0,
\eeq
and similar ones apply for $\phi^{\prime}$. Equivalently, the 5D propagator is defined by its KK decomposition
\beq
D_{\mathbb{A}}(\phi,\phi^{\prime};\hat{p})=\sum_{n}\frac{\chi_n^{\mathbb{A}}(\phi)\chi_n^{\mathbb{A}}(\phi^{\prime})}{\hat{p}^2-(m_n^{\mathbb{A}})^2} =D_{\mathbb{A}}^{(0)}(\phi,\phi^{\prime};\hat{p})+\tilde{D}_{\mathbb{A}}(\phi,\phi^{\prime};\hat{p}) ,
\label{eq:5dprop}
\eeq
where in the last step we have separated the propagator into the contribution coming from the zero mode (massless before electroweak symmetry breaking) and the one from the rest of the tower. This is useful since only $\tilde{D}_{\mathbb{A}}(\phi,\phi^{\prime};\hat{p})$, evaluated at zero-momentum $\hat{p}=0$ and in the limit $\tilde{m}_\mathbb{A}\to 0$, is relevant for EWPT at the level of dimension-6 operators. It reads
\begin{equation}
\begin{aligned}
	\tilde{D}_{\mathbb{A}}(\phi,\phi^{\prime};0) = \frac{r^2}{2} & \left[ \int_0^{\phi_>}d\phi_1 e^{2\sigma(\phi_1)} - \frac{1}{\pi} \left( \int_0^{\phi} +\int_0^{\phi^{\prime}}\right)d\phi_1e^{2\sigma(\phi_1)}\phi_1   \right. \\
	& \; \left.- \int_0^\pi d\phi_1 e^{2\sigma(\phi_1)}\left(1-\frac{\phi_1}{\pi}\right)^2 \right]+\mathcal{O}\left(\frac{\tilde{m}_\mathbb{A}^2}{M_{\rm KK}^2}\right), \label{eq:gluon_prop}
\end{aligned}
\end{equation}
where  $\phi_>=\max(\phi,\phi^{\prime})$.  Using this expression for the 5D propagator we get
\begin{equation}
	\frac{G_F}{\sqrt{2}} = \frac{1}{2v_h^2} \left[ 1+ \pi r^2\tilde{m}_{W}^2 \int_0^\pi d\phi_1 e^{2\sigma(\phi_1)} +\mathcal{O}\left(\frac{v^4_h}{M_{\rm KK}^4}\right)\right],
	\label{eq:gf}
\end{equation}
as well as the following expressions for the oblique parameters $T$ and $S$ \cite{Barbieri:2004qk,Davoudiasl:2009cd,Cabrer:2011fb}
\begin{equation}
	S= 8  v_h^2 \pi^2 r^2 \int_0^{\pi}d\phi_1e^{2\sigma(\phi_1)}\left(1-\frac{\phi_1}{\pi}\right)\,, \qquad \qquad  T=\frac{ v_{h}^2 \pi^2r^2 }{c_w^2 } \int_0^{\pi}d\phi_1 e^{2\sigma(\phi_1)}\,.		\label{eq:STparam}
\end{equation}
Note that at tree level, the contribution to the $S$ and $T$ parameters only depend on the metric and on the radius of extra dimension~$r$.

\subsection{Custodial case}
\label{sec:custodial}

One possibility of reducing the tree-level contribution to the $T$ parameter \cite{Sikivie:1980hm} and making the  $Z\bar b_Lb_L$ coupling more SM-like \cite{Agashe:2006at} is to enlarge the bulk gauge symmetry to include the custodial group $SU(2)_L\times SU(2)_R$. In particular, one can assume an $SU(2)_L\times SU(2)_R\times U(1)_X$ bulk gauge symmetry, broken by the UV boundary conditions to $SU(2)_L\times U(1)_Y$ and by the VEV of the Higgs to $SU(2)_V\times U(1)_X$. In this case, one obtains for $\mathbb{A}=Z,W^\pm$
\begin{equation} \hspace*{-0.2cm}
m_\mathbb{A}^2 = \tilde{m}_\mathbb{A}^2 \left[ 1 - \dfrac{r^2\tilde{m}_\mathbb{A}^2}{\pi}\int_0^{\pi}\!d{\phi_1} e^{2\sigma(\phi_1)} \left(\phi_1^2 + \pi^2 \rho_\mathbb{A}\right) + \mathcal{O} \left(\frac{v^4_h}{M_{\rm KK}^4}\right) \right],
\end{equation}
where now
\begin{equation}
	\tilde{m}_{W}^2=\frac{g_{5L}^2 }{2\pi r} \frac{v_{h}^2}{4},\qquad 	\tilde{m}_{Z}^2=\frac{(g_{5L}^2+g_{5Y}^2)}{2\pi r}\frac{ v_{h}^2}{4}, 
\end{equation}
and we have defined for convenience $\rho_W =t_W^2$ and $\rho_Z=t_W^2  c_{w}^2 c_\theta^2$ (see Appendix~\ref{app:gauge} for more details). 
The different mixing angles are defined as 
\begin{equation}
		t_W=\frac{g_{5R}}{g_{5L}},	\qquad t_\theta=\frac{g_{5X}}{g_{5R}}, \qquad t_w=\frac{g_{5Y}}{g_{5L}},
\end{equation}
with 
\begin{equation}
 \qquad g_{5Y}=\frac{g_{5R}g_{5X}}{\sqrt{g_{5R}^2+g_{5X}^2}}.
\end{equation}
One also obtains 
\begin{equation}
\frac{G_F}{\sqrt{2}}=\frac{1}{2v_h^2} \left[ 1+\pi r^2\tilde{m}_{W}^2  \left( 1 + \rho_W\right)\int_0^\pi\!d \phi_1 e^{2\sigma(\phi_1)}   + \mathcal{O}\left(\frac{v^4_h}{M_{\rm KK}^4}\right)\right],
\end{equation}
whereas the $T$ parameter now vanishes and the $S$ parameter remains unchanged as in \eqref{eq:STparam}.

\begin{figure}[t!]
\begin{center}
\includegraphics[scale=0.5]{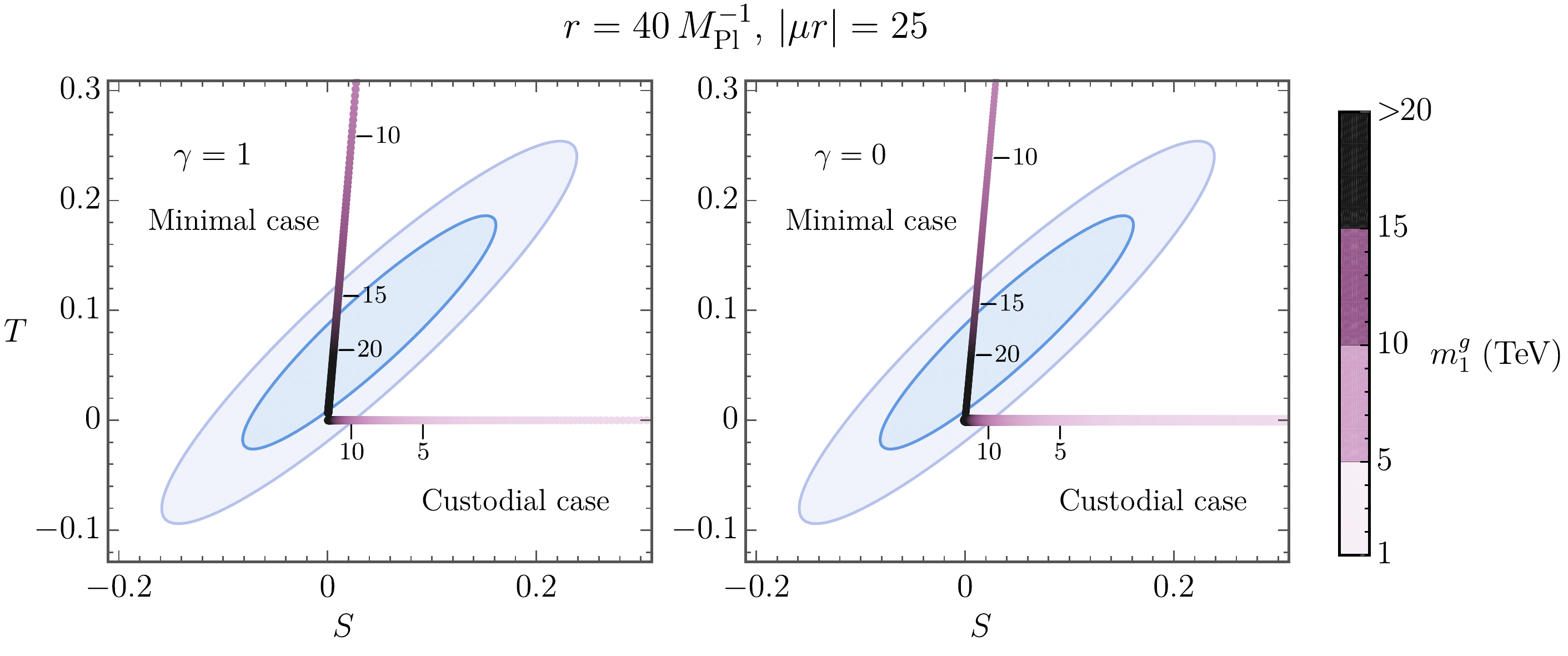}
\medskip
\caption{Values of the $S$ and $T$ parameters for the scan performed on the mass of the first KK gluon $m_1^g$ for $\gamma = 1$ (strong backreaction) and 0 (no backreaction limit). The color bar shows the values of the mass of the first KK gluon $m_1^g$ in TeV. The blue contours show the 68\% and 95\% CL regions in the $S-T$ plane allowed by EWPT~\cite{Haller:2018nnx}. For the left panel we have fixed the parameters to be $\abs{\mu r} = 25$ and $\gamma=1$. The right panel shows the no backreaction limit, i.e. $\gamma=0$, for the same value of $\abs{\mu r}$. The horizontal $T=0$ points refer to the custodial case, whereas vertically inclined points are for the minimal case. Some of the scan points are marked with their corresponding $m^g_1$ values in TeV.} \label{fig:scan_gluon}
\end{center}
\end{figure}

\subsection{Numerical study} \label{sec:EWPOs_NS}

We now derive the constraints implied by the determination of the $S$ and $T$ parameters on the KK mass scale for both the minimal and the custodial RS models, including the backreaction on the metric from the VEV of the bulk scalar field. In figure~\ref{fig:scan_gluon} we show contours of 68\% (blue) and 95\% CL (light blue) allowed by the $S$ and $T$ parameters, as reported by Gfitter group~\cite{Haller:2018nnx}. As mentioned earlier, we use the mass of the first KK gluon resonance, $m_1^g$, to display the bounds, where $m_1^g$ is obtained from the solution of the gauge boson equation of motion~\eqref{eq:gauge_eom} with Neumann boundary conditions. We perform a numerical scan, where for given values of $|\mu r|$, $kr$ and $\gamma$ we determine the KK gluon mass, as well as the warp function and the VEV of the bulk scalar from the coupled equations \eqref{eomMVEV} and \eqref{eomMVEV2}. We then obtain $S$ and $T$ from~\eqref{eq:STparam} and the corresponding ones for the custodial case.  In figure~\ref{fig:scan_gluon}, the colored legend shows the values of the  first KK-gluon mass. The size of the extra dimension is fixed to $r=40 \mpl^{-1}$. For the left panel we have fixed the parameters to be $\abs{\mu r} = 25$ and $\gamma=1$, whereas the right panel corresponds to the case where no backreaction is considered, i.e.\ $\gamma=0$. Comparing the two plots one sees that backreaction only has a minor impact on the numerical results. Also, the results are rather insensitive to changes in $|\mu r|$, as one can infere from figure~\ref{fig:background_sols_BR}.

We collect the EWPT 95\% CL bounds on the lightest KK-gluon mass $m_1^g$ in table~\ref{tab:g_mass}. One observes that regardless of the model parameters the bounds on $m_1^g$ are rather stringent and out of the reach of LHC. Note that custodial versions of the RS model are favored over models with minimal particle content in that they allow for lighter KK resonances, minimal versions of the RS model are favored by Higgs phenomenology~\cite{Goertz:2011hj, Carena:2012fk, Malm:2013jia}. We should emphasize however that our analysis of EWPT bounds was based on tree-level calculations. In some regions of parameters space, positive one-loop corrections to the $T$ parameter could help to eliminate the bounds. For the custodial RS model, this has been studied in~\cite{Carena:2006bn, Carena:2007ua}.  The backreaction of the scalar VEV on the metric only has a minor impact on the bounds on $m_1^g$. Note, however, that this would not be true if one considers the ``unphysical'' KK mass scale as defined in terms of the derivative of the metric in~\eqref{defMKK}, which varies strongly with $\gamma$.

\begin{table}[t!] \centering \renewcommand{\arraystretch}{1.2}
\begin{tabular}{ c c c c c }
\hline
$\gamma$  & $m_1^g$ (TeV)\,--\,MC & $m_1^g$ (TeV)\,--\,CC & $\mkk$ (TeV)\,--\,MC  \\
\hline
$1$                     & 14.3  & 9.6  & 21.5 \\
$0.5$					& 14.0	& 9.6	& 12.9 \\
$0.1$               & 13.9  & 9.6  & 7.0   \\
$0.01$               & 13.8  & 9.5  & 5.8   \\
$0$ (no backreaction limit)    & 13.6  & 9.5  & 5.6   \\ \hline
\end{tabular}
\medskip
	\caption{Mass of the lightest KK gluon satisfying the constraints from EWPT at 95\% CL for the minimal case (MC) and the custodial case (CC). The parameter $\gamma$, which is varied between $1$ and $0$, determines the strength of the backreaction. We fixed $\abs{\mu r}=25$ and $r=40 M_{\text{Pl}}^{-1}$. The last column shows $\mkk$ as defined in~\eqref{defMKK} for the minimal case.}
	\vspace{0.5cm}
  \label{tab:g_mass}
\end{table}

\section{Kaluza-Klein excitations of the bulk scalar field} \label{sec:scalar}

\begin{figure}[t!]
\begin{center} \hspace*{-0.9cm}
\includegraphics[scale=0.55]{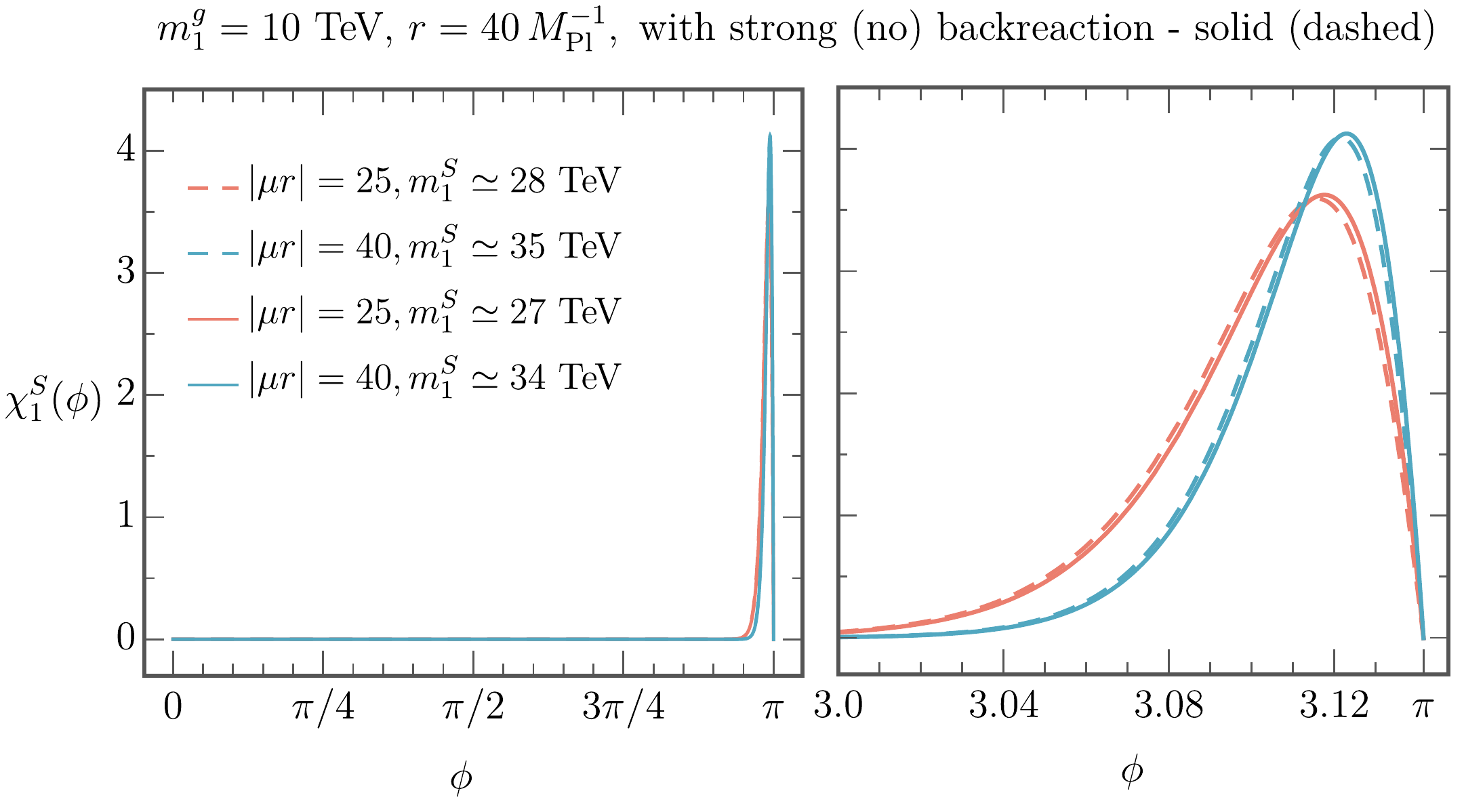} 
	\caption{Profiles of the first KK resonance of the bulk scalar field for $\abs{\mu r} =$ 25 (red) and 40 (blue), for strong (solid lines) and no (dashed lines) backreaction on the metric for $\gamma = 0.5$. The different mass eigenvalues are shown in the legend. In all cases we have fixed $m_1^g=10\tev$.} \label{FigSKK}
\end{center}
\end{figure}

The profiles of the scalar KK resonances can be computed by inserting the KK decomposition
\beq
S(x,\phi) = \sum_{n=1}^\infty S_{n}(x) \,  \chi_{n}^{S}(\phi)
\eeq
into the action~\eqref{S5daction} and keeping the quadratic terms in the field. This allows us to obtain the following equation of motion:
\beq
\left[ \partial_{\phi}^2 - 4 \, \sigma'(\phi) \,  \partial_{\phi} - \abs{\mu r}^2 \left( 3 v^2 (\phi) -1 \right) + r^2(m_{n}^{S})^2  e^{2 \sigma(\phi)} \right] e^\sigma \chi_{n}^{S} (\phi)=0 . \label{SKKphi}
\eeq
The normalization of the scalar profiles is given by
\beq
2  \int_{0}^\pi d \phi \; \chi_{n}^{S\dagger} (\phi) \, \chi_{m}^{S} (\phi) = \delta_{mn} \,.
\eeq
In figure~\ref{FigSKK} we show the numerical solutions of the lightest odd scalar KK modes for strong (solid) and no (dashed) backreaction for the input values $\abs{\mu r} = 25$ (red) and $\abs{\mu r} = 40$ (blue). We have fixed the mass of first KK gluon $m^g_1 = 10\tev$, whereas the corresponding eigenvalue (mass) of KK scalar mode is also given. Note that the  Dirichlet boundary conditions for the scalar field are incompatible with the existence of a zero-mode solution. We find the first KK excitation of the scalar field to have a mass of order 30 TeV for the input parameters leading to $m^g_1 = 10\tev$. This means, in particular, that the masses of the scalar KK resonances are out of the reach of the LHC and that the fermionic and vector KK resonances would be the first to be found in direct searches.

Even if our estimates for the  mass of the lowest-lying scalar resonances are beyond the kinematic reach of the LHC, it is illustrative to investigate which are the main decay channels. Therefore, we computed some of the relevant decay modes for the first KK mode of the bulk scalar, $S \equiv S^1$, and show the different branching ratios (BR) in table~\ref{tab:BR_S}. We  work in the massless quark limit $v_{h}\to 0$ and use the notation where $q$ stands for the $(u,d,c,s,b)$ quarks, while $Q$  denotes the KK excitations of all SM fermion chiral fields besides $t_R$, whose KK excitation is denoted by $T$. This is justified by the fact that, for the cases at hand, all the KK resonances besides $T$ share approximately the same mass $m_Q\approx 10~\tev$.  Remember that the tree-level decay into two SM quarks is forbidden in the massless quarks limit, i.e.\ when $v_h/\mkk \to 0$, as in this case there is no mixing between the zero modes and KK fermions.

We find the preferred decay channel to be the one where $S$ decays into two KK resonances, with BR($S\to QQ) \sim 80-90 \%$. Decays into one SM quark and the KK quark have significantly smaller branching rations. The decay channel $S \to TT$ is open for $\gamma = 0$ and closed for $\gamma = 0.5$, where $2 m_T > m_S$. For the decay into gluons, $S\to gg $, we have estimated the branching ratios by summing over the first few KK fermions in the loop. As one can see from the table, this mode has a small branching ratio of approximately $1 \%$. In principle, the decay channel $S \to GG$, with $G$ being the first KK excitation of the gluon, is kinematically open and should be taken into account. While  its calculation is highly non trivial we expect a branching ratio not much larger than that for $S\to gg$, which is already pretty small.  We also neglect electroweak loop processes  like $S\to \gamma\gamma, S\to W^+W^-$, $S\to ZZ$, $S\to \gamma Z$, and similar ones involving their KK resonances. We find the total decay width of the scalar to be $\Gamma/m_S \sim (2 - 4) \cdot 10^{-4}$ for the different cases presented in  table~\ref{tab:BR_S}, due to the small couplings of the scalar to the different fermions. We thus conclude that the new scalar is likely to be a very narrow resonance, with a width of order a few GeV, despite the large number of decay channels present. 

The discovery of such particle, even at a possible $100\,\tev$ collider~\cite{Benedikt:2018csr}, seems extremely challenging. We have checked that the leading-order cross-section $\sigma(gg\to S)$ for a center-of-mass energy $\sqrt{s}=100\,\tev$ and the PDF set 
\texttt{CT14nlo} is of the order of $\mathcal{O}(10^{-7})~$pb. Associated $Q$ production  $pp\to S Q$ via the $t$-channel exchange of a $Q$ resonance and the subsequent $S\to QQ$ decay seems the more promising channel to discover such an elusive resonance. The required collider study to precisely assess this is, however, beyond the scope of this paper.   

\begin{table}[t!]\renewcommand{\arraystretch}{1.2} \hspace*{-.75cm}
	\begin{center}
\begin{tabular}{ c |c c c c c c  c c}
\hline 
	$(|\mu r|,\gamma)$&  $ S\to qQ$	& $ S\to t Q $ & $S\to tT$ & $S\to QQ$ &  $S\to TT$ & $S\to gg$ &\!\! $m_S$\,(TeV)\!\!&\!\! $m_T$\,(TeV)\!\!\\
\hline
	(25, 0)	& 0.07  &0.04 	&0.01  & 0.86	 & $2\cdot 10^{-5}$ & 0.011 & 28&14\\
	(40, 0) 	& 0.06  &0.03 	&0.01  	& 0.89 & $7\cdot 10^{-4}$ & 0.008 & 35&14\\
	(25, 0.5)	& 0.16  &0.05 	&0.03 	& 0.75	 & 0 & 0.009 & 27&20\\
	(40, 0.5)	& 0.06 	&0.03 	&0.12 	& 0.77	 & 0 & 0.007 & 34& 22\\ \hline
\end{tabular}
\medskip
	\caption{Main branching ratios for the first KK resonance of the $\mathbb{Z}_2$-odd scalar, $S$, for all cases under consideration. Last two columns show the values of $m_S$ and $m_T$ while $m_Q\approx 10\,\tev$ for all the cases shown, see main text for more details.}
  \label{tab:BR_S}
\end{center}
\end{table}

\section{Bulk fermion masses and profiles}
\label{sec:fermions}

As previously mentioned, one of the main motivations behind this work is to provide a dynamical mechanism to generate fermion bulk masses and therefore to solve the flavor puzzle.  The bulk fermion bilinear $\bar\Psi\Psi$ has to be odd under $\phi\to-\phi$ in order to respect the $\mathbb{Z}_2$ symmetry of the model. Therefore, one tantalizing possibility is that these masses arise via  a bulk ``Higgs mechanism'' through the coupling to $\mathbb{Z}_2$-odd scalar field $\Sigma$ which acquires a VEV along the extra dimension $\omega(\phi)$. Contrary to the usual {\it ad-hoc} approach, where the bulk mass parameters are multiplied by $\sgn(\phi)$ to make the 5D mass term of the Lagrangian  $\mathbb{Z}_2$ invariant, now both the profile along the extra dimension and the parity of the VEV are a byproduct of the parity assignment and the dynamics of the 5D fields. In this section, we present this mechanism  of generating bulk fermion masses for a single fermion $\Psi$, singlet under the gauge group. We will generalize it to the three-generation case and the SM gauge group in the next section. 

Considering the Yukawa couplings with the bulk scalar, the fermion action can be written as~\cite{Grossman:1999ra, Gherghetta:2000qt}
\begin{equation} \hspace*{-0.4cm}
S_{\rm ferm}^{\rm (5D)} = 2 \int d^4x \! \int^\pi_{0} d \phi \sqrt{g} 
	\left\{ E_a^N \left[\dfrac{i}{2} \bar{\Psi}\, \Gamma^a (\partial_{N}-\overleftarrow{\partial}\!\!_{N}) \Psi + \frac{\omega_{bcN}}{8}\,\bar\Psi \{\Gamma^a,\sigma^{bc}\} \Psi \right] - \mathcal{Y} \bar \Psi \Sigma \Psi \right\} \,, \label{actionfer}
\end{equation}
where $\Gamma_{A} = (\gamma_{\mu} , i \gamma_{5})$ are the 5-dimensional Dirac matrices, $E_a^N=\mbox{diag}(e^\sigma \delta^{\mu}_{a},1/r)$ is the inverse f\"unfbein, and $\omega_{bcN}$ is the spin connection. In our notation, upper-case Roman indices $(M,N,\ldots)$ are used for objects defined on the curved spacetime, whereas lower-case Roman indices $(a,b,\ldots)$ correspond to objects defined in the tangent flat spacetime. The gamma matrices $\Gamma_{a}$ provide a 4D irreducible representation of the 5D Clifford algebra in flat space, i.e., $\{\Gamma_a,\Gamma_b\}=2\eta_{ab}$. Since the warped metric~\eqref{metric} is diagonal, the only non-vanishing entries of the spin connection are those with $b\!=A$ or $c=A$, giving no contribution to the action in (\ref{actionfer})~\cite{Grossman:1999ra}. The last term in the action is the 5D Yukawa interaction of the fermion $\Psi$ with the odd scalar $\Sigma $ with Yukawa coupling $\mathcal{Y}$. The mass dimensions of the fields and couplings in the above action are $[\Psi] = 2$, $[\Sigma] = 3/2$, and $[\mathcal{Y}] = -1/2$.

After the scalar field gets its VEV, the above action for the fermion field can be rewritten as
\begin{equation} \hspace*{-0.4cm}
S_{\text{ferm}}^{\text{(5D)}} = 2 \int d^4x  \int^\pi_{0} d \phi \sqrt{g} 
	\left\{ E_a^N \left[\dfrac{i}{2} \bar{\Psi} \Gamma^a (\partial_{N}-\overleftarrow{\partial}\!\!_{N}) \Psi \right] -   \omega(\phi) \mathcal{Y}  \bar\Psi \Psi\right\} +S_{\rm ferm}^{\rm int}\,, \label{actionfer2}
\end{equation}
where $S_{\rm ferm}^{\rm int}$ is the fermion interaction part with the odd scalar fluctuation $S(x,\phi)$. 
We KK-decompose the bulk fermions into the 4D chiral fermion modes as in~\cite{Grossman:1999ra},
\begin{equation}
	\Psi (x,\phi) =  \sum_{\substack{n=0 \\ A=L, R}}^\infty \psi^{A}_{n} (x) \dfrac{e^{2\sigma(\phi)}}{\sqrt{r}} f_n^{A}(\phi)\,,\label{eq:fermkk}
\end{equation}
where $\psi^{L,R}_{n} (x)$ are the left- and right-handed 4D chiral fermions with their corresponding $\phi$-dependent profiles $f_n^{L,R}(\phi)$. The prefactor $e^{2\sigma(\phi)}/\sqrt{r}$ is introduced to ensure the canonical normalization of the KK modes and to make the wave-functions $f_n^{L,R}(\phi)$ dimensionless. 
As a consequence of the above KK decomposition, the $f(\phi)$-profiles satisfy the orthogonality condition
\begin{equation}
 2 \int_{0}^\pi d \phi \,e^{\sigma(\phi)} f_m^{L,R*}(\phi)\, f_n^{L,R}(\phi) = \delta_{mn} \,.
\label{Eq:normferm}
\end{equation}
By use of the variational principle and integrating by parts in the action~\eqref{actionfer2}, we obtain the equation of motion for the fermion profiles,
\begin{align}\label{eqs1}
	\left[ \pm\frac{1}{r}\,\partial_\phi - \mathcal{Y} \ome \right]  f_n^{L,R}(\phi) = &- m_n \,e^{\sigma(\phi)} f_n^{R,L}(\phi) \,,
\end{align}
along with boundary condition
\beq
	e^{-4 \sigma} \delta \bar{\Psi} \gamma_{5} \Psi \Big\vert_{\phi=0,\pi} =  \; 0. \label{eqs1b}
\eeq
The plus (minus) sign on the left-hand side of~\eqref{Eq:normferm} refers to $f_n^{L}~(f_n^{R})$. The masses of the KK resonances are denoted by $m_{n}$. It is instructive to introduce a dimensionless quantity
\begin{equation} 
	c \equiv  \mathcal{Y} \sqrt{\dfrac{6}{\lambda}} \, \dfrac{\abs{\mu r}}{ \sigma'(\pi)} \label{eq:c}
\end{equation}
such that \eqref{eqs1} can be recast in the dimensionless form
\beq \label{eq:eqsferm2}
 \Big[\pm\partial_\phi - c \, \sigma'(\pi) \, v(\phi) \Big] f_n^{L,R}(\phi) = - r m_n e^{\sigma(\phi)} f_n^{R,L}(\phi) \,, 
 \eeq
 where we have replaced $ \omega(\phi)$ with $ v(\phi)$ as in~\eqref{normfact1}.  In the case where $v(\phi)$ is replaced by $\sgn(\phi)$, $c$ corresponds to the conventional dimensionless bulk mass parameter used e.g.\ in \cite{Casagrande:2008hr}.

The boundary conditions in~\eqref{eqs1b} guarantees either $f_m^{L*}(0)\, f_n^R(0) = 0$, or $ f_m^{L*}(\pi)\, f_n^R(\pi)=0$, i.e.\ for a given field and assuming the same $\mathbb{Z}_{2}$ behavior at both branes, either the left-handed or the right-handed component of the field should be $\mathbb{Z}_{2}$-odd. That allows us to define a $\mathbb{Z}_{2}$-parity on the orbifold to be given by $\pm \gamma_{5}$. The zero-mode profiles are obtained by solving~\eqref{eqs1} with $m_n=0$. The equation then determines the $\mathbb{Z}_2$-even profile functions, while the $\mathbb{Z}_2$-odd functions vanish.

\subsection{Fermion zero modes}

\begin{figure}[t!]
\begin{center}
\hspace*{-0.8cm}
\includegraphics[scale=0.53]{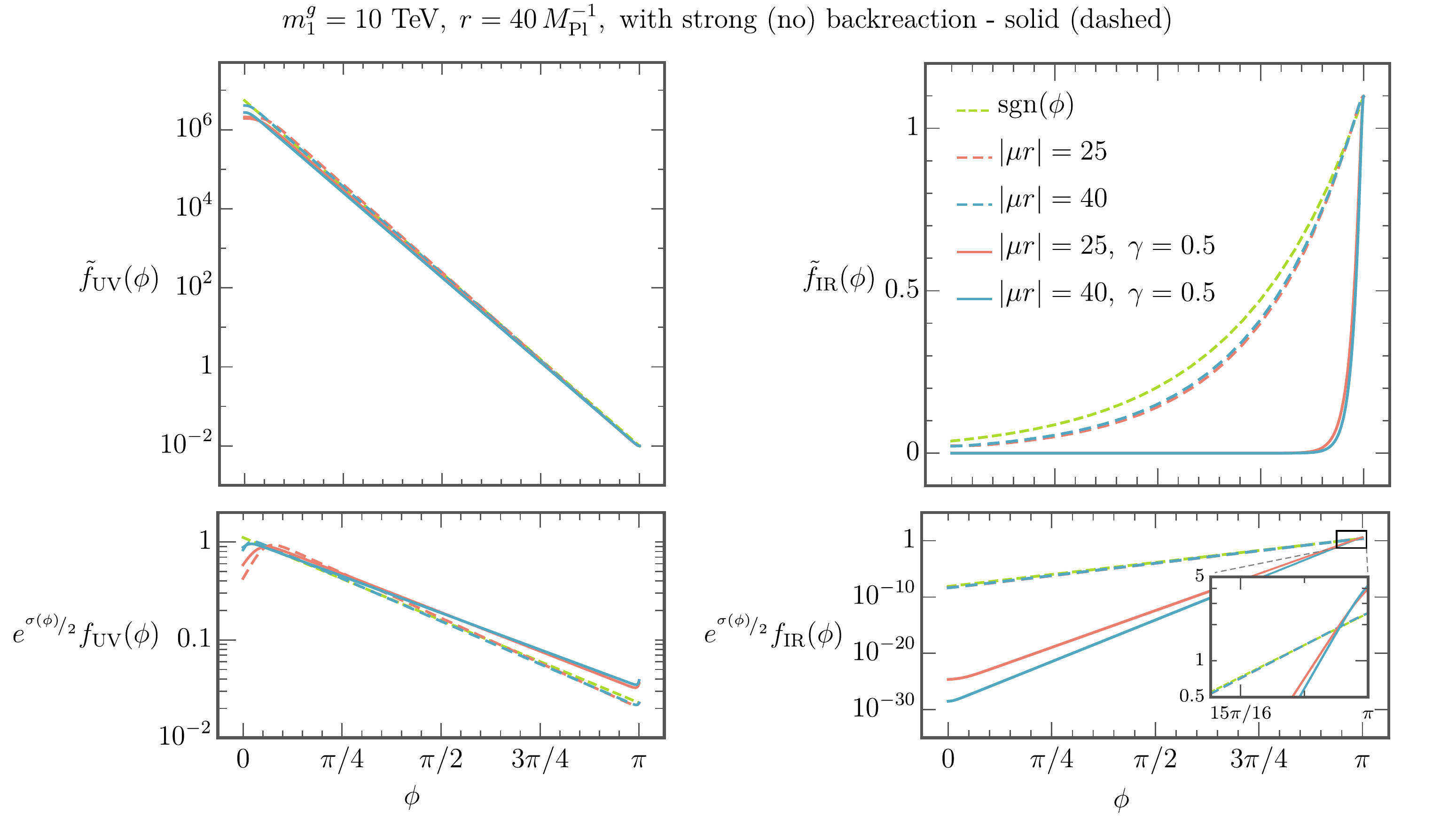}
	\caption{Examples of profiles of fermion zero modes obtained in the original RS model (green), i.e.\ using a $\sgn(\phi)$ function in front of the bulk masses, and in the model where the bulk masses are generated dynamically through the VEV of a scalar field, considering both strong and no backreaction (solid and dashed lines, respectively). We use the background solutions shown in figure \ref{fig:background_sols_BR}. 
	The left~(right) panels show profiles which are localized near the UV~(IR) brane. In each case, we adjust the dimensionless couplings $c$ such that the values of the profiles on the IR brane are the same in all cases, with $F(c)=\tilde{f}(\pi)=0.01$ and $1.1$ for the left  and right panels, respectively. 
	In the upper panels we show the rescaled profiles $\tilde{f}(\phi)$ defined in~\eqref{eq:ferm_norm_yukawas}, whereas the lower panels display the ``properly normalized'' solutions $e^{\sigma(\phi)/2}f(\phi)$, see text for more details.  } 
\label{fig:f0modes}
\end{center}
\end{figure}

The fermion profiles are now given by~\eqref{eq:eqsferm2} and need to be obtained numerically. From \eqref{eq:eqsferm2} the expression for the zero-mode profiles read:
\begin{equation}
f_0^{L,R}(\phi) = f_0^{L,R}(0) \, \exp \left( \pm c\, \sigma'(\pi)  \int^\phi_{0}  v (z) \, dz \right) . \label{eq:ferm0modes}
\end{equation}
As usual, the chirality chosen to vanish for a given fermion field at one of the branes, e.g.\ $f^R_{0} (0) = 0$ or $f^L_{0} (0)= 0$, will not have a zero mode, since the differential equation along with the boundary conditions from~\eqref{eq:eqsferm2} cannot be satisfied for a non-trivial solution.

In figure~\ref{fig:f0modes} we compare the numerical solutions for the fermion zero modes computed using the background solutions shown in figure~\ref{fig:background_sols_BR}, including strong and no backreaction effects on the metric. In each case, we fixed the value of the dimensionless 5D mass parameter $c$ such that the fermion profiles have the same value on the IR brane. This  will be particularly relevant in next section, when we study the impact on the flavor structure of the model arising from the non-trivial VEV profile $v(\phi)$ and the modification of the metric $\sigma(\phi)$, since fixed values of the fermion profiles on the IR brane correspond to fixed quark masses and mixing angles for a given 5D Yukawa matrix. In particular, we represent in the upper panels the rescaled profiles
\begin{equation}
	\tilde{f}(\phi) \equiv \sqrt{\frac{2}{\sigma'(\pi)}}e^{\sigma(\pi)/2}\, f_0^{L,R} (\phi)  \, ,
\label{eq:ferm_norm_yukawas}
\end{equation}
defined in such a way that their values at the IR brane, $\tilde{f}(\pi)\equiv F(c)$, are the factors responsible for the exponential hierarchies expected in the 4D effective Yukawa matrices, obtained after weighting the different entries of the anarchic 5D Yukawa couplings with them, see (\ref{eq:effyuk}) below.  On the other hand, the lower panels represent the ``properly normalized'' fermion profiles, $e^{\sigma(\phi)/2}f(\phi)$, that have norm equal to one when using $\|g\|^2=\int_{-\pi}^{\pi} g(\phi)^2d\phi$ . Such profiles are useful to visualize the changes on flavor observables,  since precisely these functions enter (convoluted with the corresponding 5D gauge  propagators) in $\Delta F=1$ and $\Delta F=2$ processes~\cite{Csaki:2008zd, Blanke:2008zb, Bauer:2008xb,Bauer:2009cf}. Since these convolutions  are very sensitive to the behavior of the functions near the IR brane, observing the change of these  profiles in the vicinity of the IR brane gives a good idea of the impact of the scenarios examined here on flavor-violating processes. 

The green dashed line in the different panels shows for comparison the profiles obtained in the conventional RS model without the bulk scalar field, where the 5D bulk masses are multiplied with $\sgn(\phi)$. We observe that in the case of no backreaction our solutions lie rather close to the green dashed line, meaning that our dynamical mechanism for generating brane-localized fermion profiles  succeeds to give profile functions which are numerically close to those of the conventional RS model. The backreaction on the metric, on the other hand, can have a significant impact on the shapes of the profiles. The phenomenological consequences of this observation will be studied later.

\subsection{Fermion KK modes}

\begin{figure}[t!]
\begin{center} \hspace*{-0.5cm}
\includegraphics[scale=0.48]{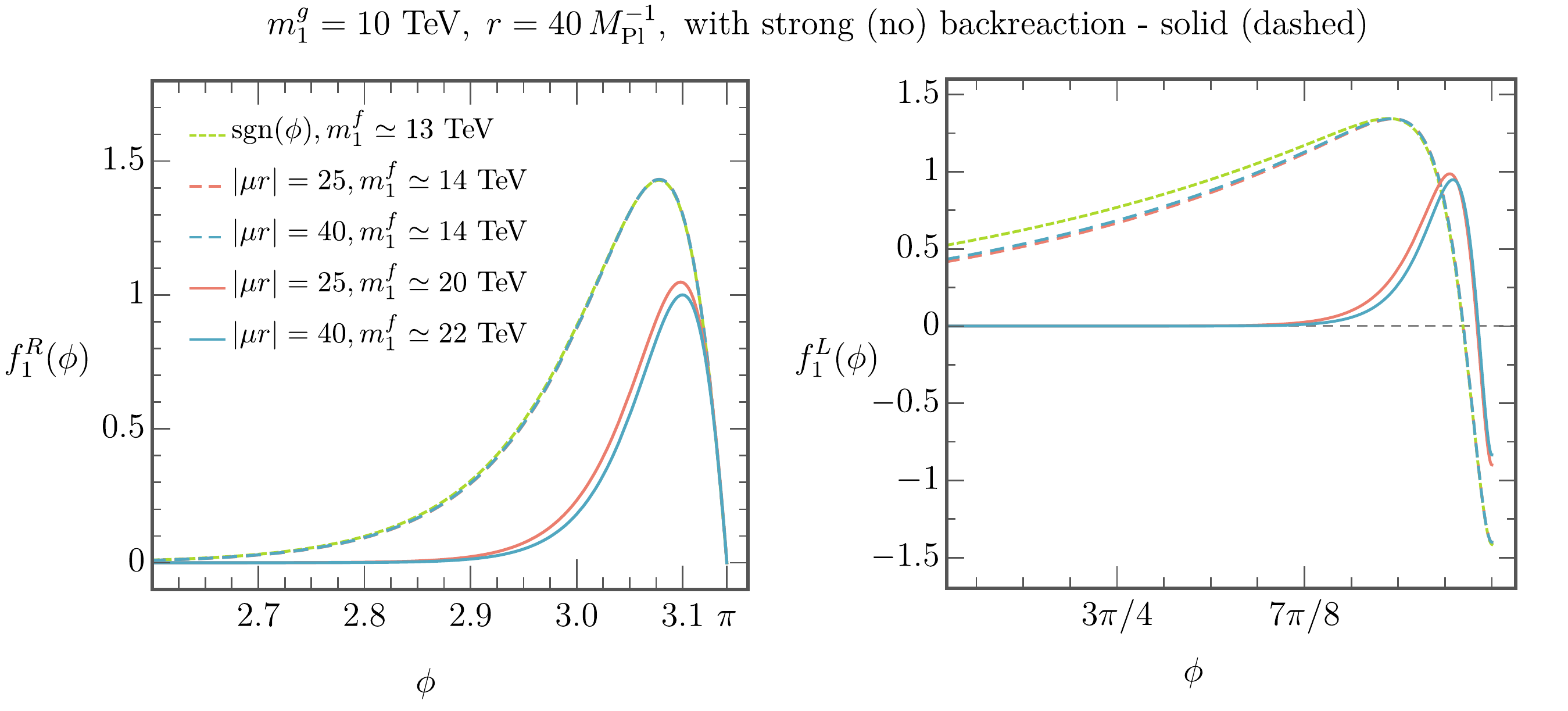} 
	\caption{Examples of profiles of the first KK resonances of a fermion with a left-handed zero mode,  obtained in the original RS model (green), i.e.\ using a $\sgn(\phi)$ function  in front of the bulk masses, and in  the model where the bulk masses are generated dynamically through the VEV of a scalar field, considering strong and no backreaction (solid and dashed lines, respectively). We use the background solutions shown in figure \ref{fig:background_sols_BR}. The left~(right) panels show the profiles for the right-handed~(left-handed) chiralities. } \label{fig:f1KK}
\end{center}
\end{figure}

For higher KK resonances the equations of motion~\eqref{eq:eqsferm2} can be decoupled by taking a derivative w.r.t.\ the fifth coordinate. This gives
\begin{equation}
\begin{split}
\Big[ \partial_{\phi}^2 - \sigma'(\phi) \,  \partial_{\phi}    & \mp c\, \sigma'(\pi) v'(\phi) + r^2 m_{n}^2 \, e^{2\sigma(\phi)} \\
&  \pm c\, \sigma'(\pi) \sigma'(\phi) v(\phi) - c^2 \, \sigma'^2(\pi) v^2 (\phi)  \Big] f_{n}^{L,R} (\phi) = 0 .
\end{split}
\end{equation}
As an example, we present in figure~\ref{fig:f1KK} numerical solutions for the right-handed and left-handed profile functions for the first KK excitation of a fermion with left-handed zero mode.  The meaning and labeling of the various curves is the same as in figure~\ref{fig:f0modes}. For each choice of parameters, we also give the value  of the corresponding KK fermion mass.

\section{A dynamical solution to the flavor puzzle}
\label{sec:pheno}

One of the major strengths of RS models is that they not only provide a solution to the hierarchy problem but also naturally explain the  large hierarchies observed in the spectrum of SM quark masses and mixing angles.\footnote{The lepton case is a bit more problematic, for the PMNS mixing matrix features large mixing angles.} In this section, we will  investigate whether this continues to be the case in scenarios where the bulk masses are generated dynamically. We will also study the implications of the dynamical approach for flavor physics. In particular, we will explore how the modifications of the fermion profiles affect the usual RS-GIM mechanism~\cite{Gherghetta:2000qt, Agashe:2004cp, Agashe:2005hk}. and study their impact on the $\Delta F=2$ observable $\epsilon_K$, measuring CP violation in $K-\bar{K}$ mixing, which typically sets the strongest flavor constraint in models with warped extra dimensions~\cite{Csaki:2008zd, Blanke:2008zb, Bauer:2008xb, Bauer:2009cf}.

The Yukawa interactions with the Higgs field, localized on the IR brane, induce a mixing between the (would-be) zero modes and the  KK excitations of different 5D fields.  These contributions are of $\mathcal{O}(v^2_h/\mkk^2)$ for both the mass eigenvalues  and the CKM mixing matrix. In order to simplify the calculations and have approximate expressions describing fermion mixings we consider the so-called ``zero-mode approximation'' (ZMA)  \cite{Grossman:1999ra,Gherghetta:2000qt,Casagrande:2008hr}, in which the fermion bulk profiles are solved with the Yukawa couplings  ``switched-off'', and these couplings are included  as a perturbation. Actually, since the overlaps of the different quark profiles with the Higgs profiles are proportional to the corresponding quark masses (up to some common multiplicative factor, see e.g.\ \cite{Casagrande:2008hr}), this approximation works best for  the light quarks, in which case the neglected terms are indeed much smaller than $\mathcal{O}(v^2_h/\mkk^2)$.

\subsection{Hierarchies of fermion masses and mixings}
\label{sec:KKferm}

We now consider three generations of 5D fermions in the bulk interacting with a scalar sector consisting of the odd scalar in the bulk and the Higgs doublet on the IR brane. If we denote by $Q$ and $q^c$ the three-component vectors in flavor space belonging to $SU(2)_L$ doublets and singlets, respectively, the quadratic terms in the 5D action can be written in the form
\begin{equation} 
\begin{split}
S_{\text{ferm, 2}}^{\text{(5D)}} = & \,\int d^4x  \int^\pi_{-\pi} d \phi \sqrt{g} 
	\left\{ \dfrac{i}{2} E_a^N \bigg[ \bar{Q} \, \Gamma^a (D_{N}-\overleftarrow{D}\!_{N}) Q + \sum_{q=u,d} \bar{q}^c \, \Gamma^a (D_{N} -\overleftarrow{D}\!_{N}) q^c \bigg] \right. \\
	&  -   \omega(\phi) \bigg[ \bar{Q}  \bm{\mathcal{Y}}_Q Q - \sum_{q=u,d}  \bar{q}^c \bm{\mathcal{Y}}_q q^c \bigg] \\
& \left. - \delta(|\phi|-\pi)\, \dfrac{\sqrt{|\hat{g_i}|}}{g} \frac{v_h \, e^{\sigma(\pi)} }{\sqrt{2}} 
     \left[ \bar u_L\,{\bm  Y}_u^{\rm (5D)}\,u_R^c
    + \bar d_L\,{\bm Y}_d^{\rm (5D)}\,d_R^c + \mbox{h.c.} \right] \right\} \,, \label{actionfer_SM}
\end{split}
\end{equation}
similar to \eqref{actionfer2}, where now the Yukawa couplings to the Higgs sector have also been introduced. Without loss of generality we work in the ``bulk-mass basis'' \cite{Casagrande:2008hr}, where $\bm{\mathcal{Y}}_{Q,q}$ denote diagonal matrices leading to real bulk masses, and ${\bm Y}_q^{\rm (5D)}$ are the 5D Yukawa matrices. Bulk Yukawa couplings can be positive or negative,\footnote{In the 5D theory, the sign of the bulk Yukawa coupling cannot be reversed by a field redefinition.} allowing for different localizations of the fermion profiles.

The 5D up-type quarks can be KK-decomposed as follows (the derivation for the down-type quarks can be trivially obtained from this one):
\begin{align}\label{KKdecomp}
	u_L(x,\phi) & = \frac{e^{2\sigma(\phi)}}{\sqrt r} \sum_n \bm{\mathcal{C}}_n^{(Q)}(\phi)\,u_L^{(n)}(x) \,, \quad
	&u_R(x,\phi) &= \frac{e^{2\sigma(\phi)}}{\sqrt r} \sum_n \bm{\mathcal{S}}_n^{(Q)}(\phi) \,u_R^{(n)}(x) \,, \\
	u^c_L(x,\phi) &= \frac{e^{2\sigma(\phi)}}{\sqrt r} \sum_n \bm{\mathcal{S}}_n^{(u)}(\phi)\,u_L^{(n)}(x) \,, 
	&u^c_R(x,\phi) &= \frac{e^{2\sigma(\phi)}}{\sqrt r} \sum_n \bm{\mathcal{C}}_n^{(u)}(\phi)\,u_R^{(n)}(x) \,, 
\end{align}
where the diagonal matrices $\bm{\mathcal{C}}_n^{(Q,u)}$ denote $\mathbb{Z}_2$-even profiles, while $\bm{\mathcal{S}}_n^{(Q,u)}$ are also diagonal matrices and correspond to odd profiles. The index $n$ labels the mass eigenstates, with fermion masses $m_n$ and spinor fields $u_{L,R}^{(n)}(x)$. The spinor fields on the left-hand side of the equations are three-component vectors in flavor space. 
The equations of motion for the different profiles in the ZMA correspond to those presented in section~\ref{sec:fermions}.

Inserting these decompositions into the action, the Yukawa interaction on the IR brane reads
\begin{equation}
	S^{\text{(4D)}} \supset \, - \int d^4x \; \frac{v_h \, e^{\sigma(\pi)} }{\sqrt{2} \, r}\left[ \bar u_L^{(n)}(x)\,\bm{\mathcal{C}}_n^{(Q)} (\pi)\,\bm{Y}_u^{\rm (5D)}\, \bm{\mathcal{C}}_m^{(u)} (\pi) \,u_R^{(m)}(x)+ \mbox{h.c.} \right] , \label{actionfer_SM_2}
\end{equation}
where it is convenient to use the fermion fields normalized as in \eqref{eq:ferm_norm_yukawas}. It is clear from \eqref{eq:ferm0modes} that, for a given pair of background solutions $\sigma(\phi)$ and $v(\phi)$, the values of the profiles of the different zero modes on the IR brane depend only on the dimensionless quantity $c$, defined in \eqref{eq:c}. Therefore, it is useful to define the functions
\begin{equation}
	F(c_{X,i}) = \left(\tilde{\bm{\mathcal{C}}}^{X}_{0} (\pi)\right)_{ii} \,,
\end{equation}
i.e., the value of the zero-mode profile of the field $X=Q,u,d$ on the IR brane, which depend only on the parameter $c_{X,i}$. Using this convention, we define the dimensionless Yukawa and the effective 4D Yukawa of the SM fields
\begin{equation}
	\bm{Y}_q^{\text{(5D)}} = \dfrac{2 \, \bm{Y}_q \, r}{\sigma'(\pi)}, \qquad (\bm{Y}_{q}^{\text{eff}})_{ij} =  F(c_{Q,i})  \left(\bm{Y}_q \right)_{ij} F(c_{q,i}),
	\label{eq:effyuk}
\end{equation}
as in \cite{Casagrande:2008hr}, where $\|\bm{Y}_q \|\sim \mathcal{O}(1)$. Once the Yukawa couplings on the IR brane are ``switched on'' a mass matrix is obtained for the SM fields. The physical masses are obtained by solving the eigenvalue equation
\begin{equation}
	\text{det} \Big(\bm{I} \, m_n^2 - \dfrac{v^2_h}{2} (\bm{Y}_{q}^{\text{eff}}) (\bm{Y}_{q}^{\text{eff}})^\dagger \Big) = 0 \,.
\end{equation}
The eigenvectors of the matrices $\bm{Y}_q^{\rm eff} \left(\bm{Y}_q^{\rm eff} \right)^\dagger$ and $\left( \bm{Y}_q^{\rm eff} \right)^\dagger \bm{Y}_q^{\rm eff}$ (with $n=1,2,3$ and $Q=U,D$, $q=u,d$) form the columns of the unitary matrices $\bm{U}_q$ and $\bm{W}_q$ appearing in the singular-value decomposition
\begin{equation}
	\bm{Y}_q^{\text{eff}} = \bm{U}_q\,\bm{\lambda}_q\,\bm{W}_q^\dagger \,,
\end{equation}
where
\begin{equation}
	\bm{\lambda}_u = \frac{\sqrt{2}}{v_h}\,\mbox{diag}(m_u,m_c,m_t) \,, \qquad 
	\bm{\lambda}_d = \frac{\sqrt{2}}{v_h}\,\mbox{diag}(m_d,m_s,m_b) \,.
\end{equation}
In this approximation, the fields are mixed because of the Yukawa interactions on the IR brane. The 5D mass eigenstates and the SM mass eigenstates are related through the matrices $\bm{U}_q$ and $\bm{W}_q$, and the CKM mixing matrix is given by
\begin{equation} \label{eq:VCKM}
 \bm{V}_{\text{CKM}} = \bm{U}_u^\dagger\, \bm{U}_d \,.
\end{equation}
We fit the bulk mass parameters $\bm{c}_X$ from the values of $F(\bm{c}_X)$ for a randomly generated Yukawa matrices $\bm{Y}_u$ and $\bm{Y}_d$. Assuming a hierarchical structure of the zero-mode profiles on the IR brane, i.e.\
\begin{equation}
\abs{F(c_{X_1})} < \abs{F(c_{X_2})} < \abs{F(c_{X_3})} \,,
\end{equation}
the system is not determined and the different $F(c_{X,i})$ parameters can be expressed in terms of a single one \cite{Casagrande:2008hr}, which we chose to be $F_2\equiv F(c_{Q_2})$.

\subsection[Impact on flavor constraints: the example of $\epsilon_K$]{\boldmath Impact on flavor constraints: the example of $\epsilon_K$}
\label{sec:kaonmixing}

We now explore how the changes in the fermion profiles caused by the non-trivial $\phi$ dependence of the VEV of the bulk scalar field (see figure \ref{fig:f0modes}) affect observables in flavor physics. Since CP violation in $K$--$\bar{K}$ mixing, as measure by the parameter $\epsK$, provides the most stringent flavor constraint in models with warped extra dimensions~\cite{Csaki:2008zd}, we restrict our study to this particular case.  We follow closely the analyses on $\epsK$ in RS models from \cite{Bauer:2008xb,Bauer:2009cf,Bauer:2011ah, Casagrande:2008hr} and adopt the following parametrization for new physics \cite{Ciuchini:1998ix}:
\beq\label{eq:Leff}
   {\cal L}_{\rm eff}^{\Delta S=2} 
   = \sum_{i=1}^5 C_i\,Q_i + \sum_{i=1}^3 \tilde C_i\,\tilde Q_i \,,
\eeq
where the operators relevant for $K$--$\bar K$ mixing are 
\begin{align}
   Q_1 &= (\bar d_L\gamma^\mu s_L)\,
    (\bar d_L\gamma_\mu s_L) \,,    \quad  \tilde Q_1  = (\bar d_R\gamma^\mu s_R)\,
    (\bar d_R\gamma_\mu s_R) \,, \hspace{-0.25cm} \\[1mm]
   Q_4 &= (\bar d_R s_L)\,(\bar d_L s_R) \,,  \qquad \quad Q_5= (\bar d_R^\alpha s_L^\beta)\,(\bar d_L^\beta s_R^\alpha) \,.
\end{align}
A summation over color indices $\alpha,\beta$ is understood. In our convention the Wilson coefficients include only the new-physics contributions, i.e.\ $C_i\equiv C_i^{\rm NP}$. For simplicity we only consider the leading contribution, arising from the tree-level exchange of KK gluons, and neglect contributions involving the exchange of other gauge bosons or scalar fields. In the ZMA, these contributions take the form 
\begin{align}\label{eq:WCsepsK}
C_1 & = 4 \pi^2\alpha_s\, r^2 \left(1 - \dfrac{1}{N_c} \right)
 (\widetilde\Delta_D)_{12}\otimes(\widetilde\Delta_D)_{12} \,, &\qquad 
 \tilde{C}_1 &= C_1\big|_{D \to d} \,, \\
C_4 & = - 4 \pi^2\alpha_s\, r^2 \, (\widetilde\Delta_D)_{12}\otimes(\widetilde\Delta_d)_{12}  \,, &\qquad 
C_5  &= - \frac{C_4}{N_c} \, , 
\end{align}
where $\alpha_s$ is the strong coupling constant and $N_c=3$ stands for the number of colors. 
The notation $(\widetilde\Delta_X)_{12}\otimes(\widetilde\Delta_Y)_{12}$ is defined as
\begin{equation} 
\begin{split}
	(\widetilde\Delta_X)_{12}\otimes(\widetilde\Delta_Y)_{12}  
	&=  (\mathcal{U}_X^\dagger)_{1 j}(\mathcal{U}_X)_{j2} (\mathcal{U}_Y^\dagger)_{1 k}(\mathcal{U}_Y)_{k 2} \\
&\quad \times \int^\pi_{0}  d \phi \int^\pi_{0}  d \phi' \int^{\phi_<}_{0} d \bar{\phi} \, e^{\sigma(\phi)} e^{\sigma(\phi')} e^{2\sigma(\bar{\phi})} \left(\tilde{\mathcal{C}}^{X}_{0,j} (\phi) \right)^2 \left(\tilde{\mathcal{C}}^{Y}_{0,k} (\phi') \right)^2 \, , \label{eq:mixing_integrals}
\end{split}
\end{equation}
where $\mathcal{U}_D=U_d$ and $\mathcal{U}_d=W_d$. Note that in this limit the Wilson coefficients relevant for computing $\epsK$ only depend on the background solutions and the fermion profiles, and therefore on the Yukawa couplings. It can be seen from \eqref{eq:gluon_prop} that only the first term in the gauge propagator appears in \eqref{eq:mixing_integrals}, since the other terms cancel due to orthogonality and unitarity of the fermion profiles and the $\mathcal{U}$-matrices, respectively. It is clear from (\ref{eq:mixing_integrals}) that the more a particular fermion has its ``properly normalized'' profile $e^{\sigma(\phi)/2}\mathcal{C}_{0}(\phi)$ shifted away from the IR brane, the smaller are its contributions to the Wilson coefficients.  Moreover, one can readily conclude from this and from the behavior of the different cases shown on the lower panels of figure~\ref{fig:f0modes} that large values of the backreaction tend to worse the GIM mechanism present in RS models, as these ``properly normalized'' profiles grow on the vicinity of the IR brane for both UV-localized and IR-localized fermions. 

In terms of the effective Lagrangian \eqref{eq:Leff}, the quantity $\epsilon_K$ is given by
\begin{equation}
\epsilon_K = \frac{-\kappa_\epsilon\,e^{i\varphi_\epsilon}} {\sqrt2\,(\Delta m_K)_{\rm exp}}\, \mbox{Im}\,\langle K^0|\,{\cal L}_{\rm eff}^{\Delta S=2}\,|\bar K^0\rangle \,.
\end{equation}
The current experimental value \cite{Tanabashi:2018oca} and SM prediction \cite{Brod:2011ty, Straub:2018kue} for $\abs{\epsK}$ are
\begin{equation}
\abs{\epsK}^{\rm exp}=(2.228\pm0.011)\times10^{-3}\, , \qquad \abs{\epsK}^{\text{SM}} = (1.81\pm0.21)\times10^{-3}\,. \label{eq:epsK_values}
\end{equation}
To get a better understanding for the different contributions playing a role in $\epsilon_K$, it is useful to note that
\begin{equation}
|\epsilon_K-\epsilon_K^{\text{SM}}|\propto \Im\left[C_1+\tilde{C}_1+213\left(C_4+\frac{C_5}{N_c}\right)\right]
	\label{eq:epskanal}
\end{equation}
for a matching scale  $\mu_{\rm NP}=15\,$TeV. 

\begin{figure}[t!]
\begin{center}
\includegraphics[scale=0.5]{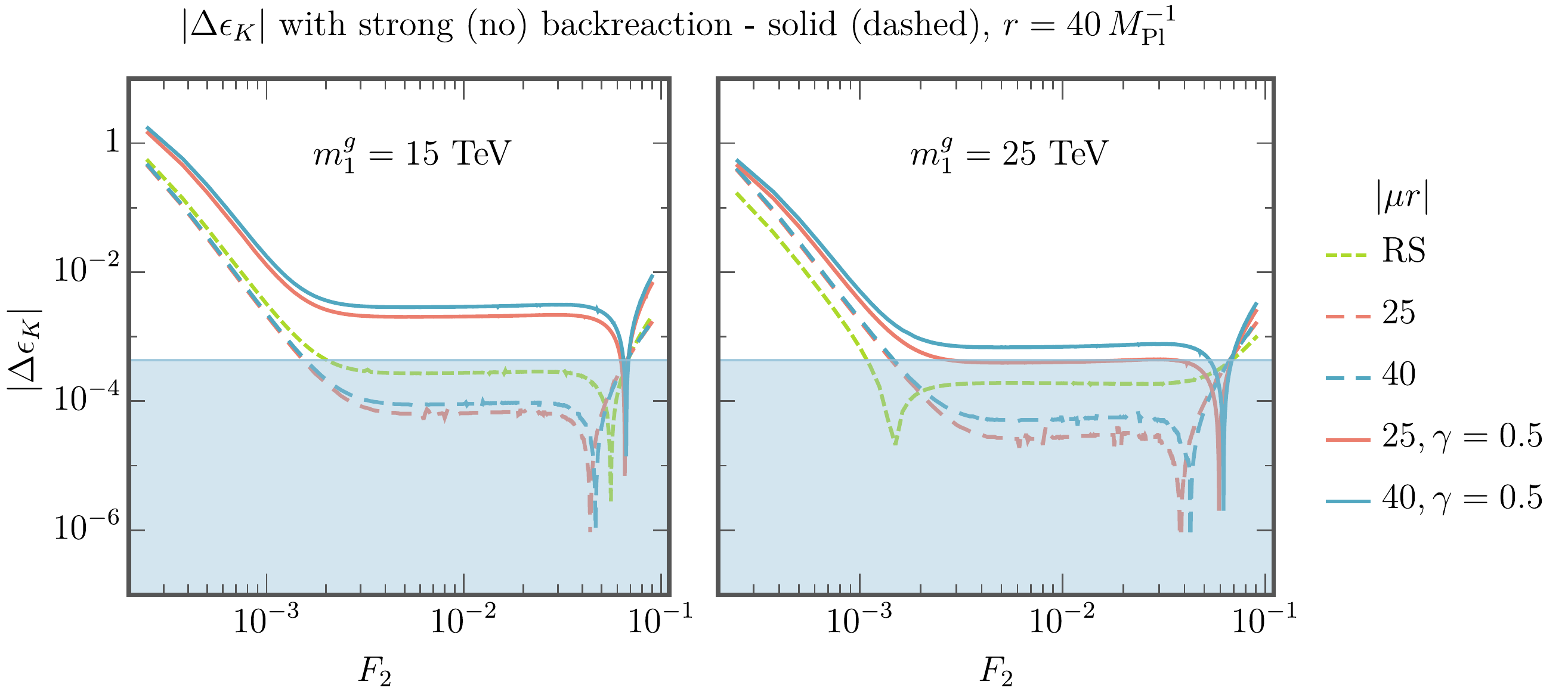}
\medskip
	\caption{$|\DepsK|$ as a function of $F_2$, obtained in the original RS model (green) and in our model, where the bulk masses are generated dynamically, for different values of $\abs{\mu r}$ and considering strong and no backreaction (solid and dashed lines, respectively). In the left (right) figure we have fixed $m^g_1=15 \tev$ ($25\tev$).} 
\label{fig:epsK}
\end{center}
\end{figure}

For our numerical analysis we generate random Yukawa matrices with complex entries of norm in the interval $[1/3,3]$ and use a $\chi^2$ distribution to select parameter sets for which the quark masses and CKM parameters are reproduced within 95\% CL. 
In figure~\ref{fig:epsK} we plot the prediction for $\DepsK\equiv |\epsilon_K|-|\epsilon_K|^{\rm exp}$ assuming different values of $\abs{\mu r}$, for strong and no backreaction, as a function of $F_2$. The left and right panels correspond to $m^g_1=15\tev$ and $25\tev$, respectively. Moreover, for comparison we also show the conventional RS case, corresponding to bulk fermion masses multiplied by a sgn($\phi$) function.
For the purposes of making this figure, we consider a benchmark point defined by the following Yukawa matrices: 
\begin{align}
	\bm{Y}_u &=
\begin{pmatrix}
    0.109 + 1.865 \, i  & & 2.000 - 1.324 \, i & & -0.706 + 1.514 \, i \\
    -2.163 - 0.615 \, i & & -0.695 - 0.483 \, i & & 2.299 - 1.604 \, i \\
    1.517 + 1.399 \, i & & -0.928 + 0.065 \, i & & 2.204 + 0.970 \, i
\end{pmatrix} 
, \\
	\bm{Y}_d &=
\begin{pmatrix}
    1.888 - 1.915 \, i & & 1.181 - 2.696 \, i & & 0.294 + 0.530 \, i \\
    1.827 - 0.057 \, i & & 2.210 - 0.413 \, i & & 0.591 + 0.951 \, i \\
	1.264 + 2.004 \, i & &  -0.829 - 1.309 \, i & &  -1.326 + 0.510 \, i
\end{pmatrix}.
	\label{eq:bench}
\end{align}
We also show with a blue band the constraint 
\beq
\DepsK\lesssim4.2\times10^{-4}\,, 
\eeq 
 which corresponds to the $2\sigma$ uncertainty of the SM prediction for $\epsK$ derived in \cite{Brod:2011ty, Straub:2018kue}. We use this value as the main uncertainty for our prediction and demand that the central value of the experimental measurement lies within that range. Note that the uncertainty of the SM prediction is one order of magnitude larger than the experimental error, as can be seen in \eqref{eq:epsK_values}. One can see from the left plot in figure~\ref{fig:epsK} (i.e.\ for $m^g_1=15\tev$) that large values of the backreaction are disfavored by the current experimental measurement of $\epsilon_K$. However, if we increase the KK scale to $m^g_1=25\tev$ (right plot in figure~\ref{fig:epsK}), then for a wide range of $F_2$ values only the strong backreaction limit for $|\mu r|=40$ remains in slight tension with the measured value of $\epsilon_K$. Note that our modified case without backreaction slightly improves the $\epsilon_K$ constraints as compared to the conventional RS case. However, as one can see from the above figure, considering a strong backreaction (with $\gamma=0.5$) leads to a larger contribution to $\epsilon_K$ and hence to a more stringent constraints on the mass of the first KK gluon mode. This is a direct consequence of the behavior of the different ``properly normalized'' profiles shown in the lower panels of figure~\ref{fig:f0modes}. For both UV and IR-localized fermions (corresponding to the lower left and right panels in figure~\ref{fig:f0modes}, respectively), the inclusion of the backreaction results in an increase of the profiles in the vicinity of the IR brane. The values of these profiles near the IR brane play a central role for flavor-violating processes, because the different vector mediators inducing flavor-changing interactions are localized near the IR brane. Thus, this results in a smaller flavor protection and hence a weakening of the RS-GIM mechanism~\cite{Gherghetta:2000qt, Agashe:2004cp, Agashe:2005hk}, in agreement with figure~\ref{fig:epsK}.  On the other hand, the shape of all curves in Figure~\ref{fig:epsK} can be easily understood  by taking into account that, for small values of $F_2$, all left-handed doublets are more and more UV localized, which makes their right-handed counterparts more and more IR localized in order to still produce the observed spectrum of fermion masses. Therefore, very small values of $F_2$ correspond to values where $\tilde{C}_1$ gets arbitrarily large to the point that can overcome the factor $213$ in front of $C_4$ in (\ref{eq:epskanal}). On the contrary, large values of $F_2$ correspond to values where exactly the opposite happens, leading to more and more IR-localized left-handed doublets and therefore to very large values of $C_1$.\footnote{Note that both limits are constrained by reproducing the top-quark mass.} For values where this arbitrary enhancement of one particular chirality ceases to occur, the dominant contributions to $\epsilon_K$ arise from $C_4$ and $C_5$, both of which are more or less constant since they involve both fermion chiralities and are therefore expected to be controlled by the different fermion masses. This is due to the fact that, as one can readily see from~\eqref{eq:effyuk}, the effective Yukawa couplings and the corresponding SM fermion masses depend on the product of the profiles of both fermion chiralities on the IR brane. In the RS case, this leads to $C_4=-N_c C_5 \sim 4\pi \alpha_s (kr\pi /\mkk^2) (2 m_s m_d /v^2 )$~\cite{Csaki:2008zd, Bauer:2009cf}. This explains the large plateau obtained for intermediate values of $F_2$ in all curves. Finally, the dips present in $\Delta \epsilon_K$ for some values of $F_2$ are the result  of a cancellation between $\Im (C_1)$ and $\Im(C_4)$, which turn out to have opposite signs for our particular choice of Yukawa matrices. 

\begin{figure}[t!]
\begin{center}
\includegraphics[scale=0.42]{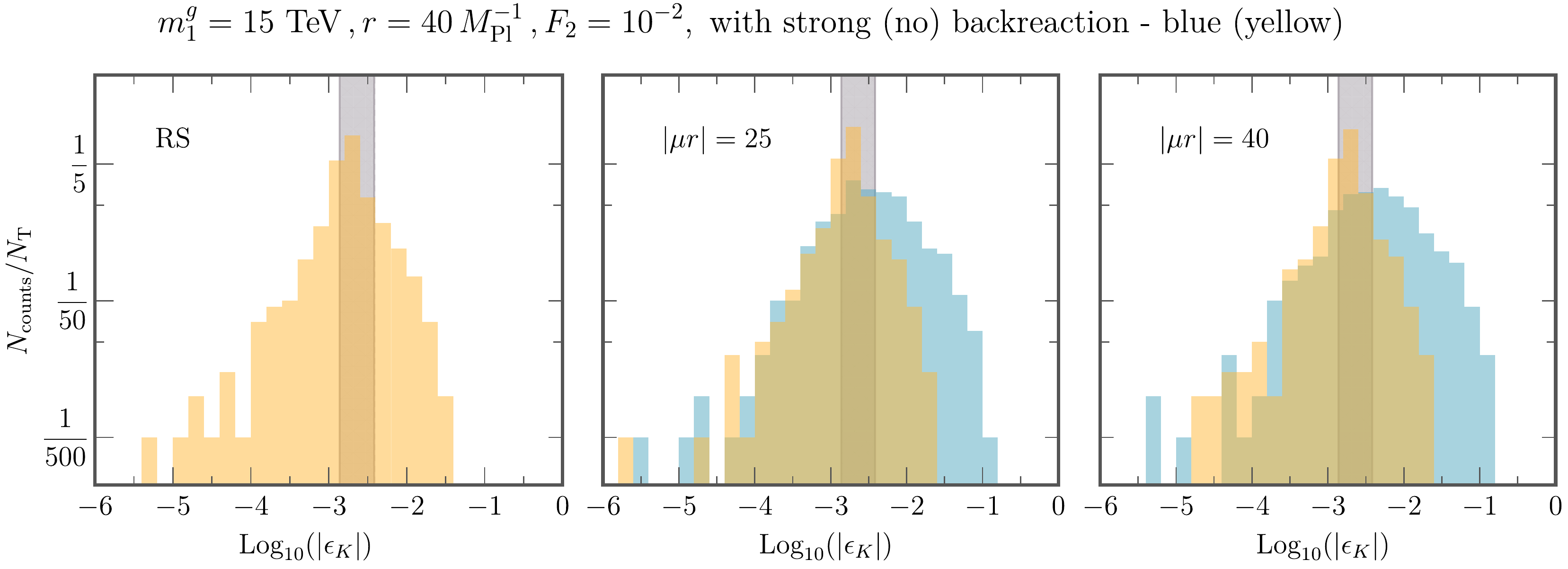}
\caption{Histogram showing the values of $\abs{\epsK}$ in the original RS model, and in our model for $\abs{\mu r}=25$ and $40$, both in the case of zero and strong backreaction (blue and yellow bins respectively). In all these figures we fix $F_2 = 10^{-2}$ and $m^g_1=15 \tev$.} 
\label{fig:hist_epsK}
\end{center}
\end{figure}

Finally, in order to have a broader picture of the flavor impact of the non-trivial $\phi$ dependence of the bulk scalar VEV and the changes in the warp factor due to the backreaction, we computed the value of $\epsK$ for a set of randomly generated Yukawa matrices. We choose a value of $F_2=0.01$, such that $C_4$ and $C_5$ give the leading contributions to $\epsilon_K$ and no large cancellations among different contributions occur. Again, we used a $\chi^2$ distribution for the generation of the Yukawa matrices. In figure~\ref{fig:hist_epsK} we show the histograms for the values of $\abs{\epsK}$ obtained from a set of $N_{T} = 500$ pairs of Yukawa matrices $\bm{Y}_u$ and $\bm{Y}_d$, for the different values of $|\mu r|$ and $\gamma$ under consideration. For generating these points we have fixed $m_1^g = 15$ TeV. It can be seen that, as we found for the benchmark point given in~\eqref{eq:bench}, a small backreaction leads to a better agreement with $|\epsK|^{\rm exp}$, whereas the strong backreaction case leads to a broader distribution and therefore to larger tension with data.

\subsection{Flavor structure of low-lying scalar KK resonances}

In extensions of the SM featuring a warped extra dimension, the presence of TeV-scale KK resonances offers new possibilities for inducing flavor-changing interactions among the fermions of the SM. The flavor-changing couplings of KK gluons and other KK gauge bosons have been explored in detail in the literature \cite{Agashe:2004ay,Agashe:2004cp,Casagrande:2008hr,Blanke:2008zb,Csaki:2008zd,Blanke:2008yr,Bauer:2009cf}. 
At low energies, they give rise to various dimension-6 interactions in the effective weak Hamiltonian, whose contributions are suppressed by two powers of the KK mass scale ($\sim 1/\mkk^2)$.

In our model, the presence of scalar KK excitations provides a new source of flavor violation, and it is interesting to study the low-energy manifestations of this effect. The couplings of the lowest-lying scalar KK resonance $S_1$ to SM quarks can be parameterized in the form
\begin{equation}
   {\cal L}_{\rm ferm} = - \sum_{q=u,d}\,\sum_{m,n}\,S_1(x) 
    \left[ g_{mn}^{(q)}\,\bar q_L^{(m)}(x)\,q_R^{(n)}(x) + \text{h.c.} \right] ,
\end{equation}
where $m,n=1,2,3$ are generation indices. Analogous couplings can be written in the lepton sector. An explicit expression for the quantities $g_{mn}^{(q)}$ has been presented in eq.~(3.26) of \cite{Bauer:2016lbe}, where a model profile $\chi_1^{S}(\phi)\propto e^{(1+\beta)kr|\phi|}$ was assumed, with a free parameter $\beta$. For our purposes, all we need to do is to replace this profile by the function $\chi_1^{S}(\phi)$ we have obtained from the solution to the equation of motion \eqref{SKKphi}. For simplicity, we will neglect the effects of backreaction in the following discussion. Using the ZMA for the SM fermion profiles, we then obtain (with $t\equiv\epsilon\, e^{kr|\phi|}$ and $\epsilon=e^{-kr\pi}$) 
\begin{equation}
\begin{aligned}
   g_{mn}^{(u)} &= \int_\epsilon^1\!dt\,\chi_1^{S}(t)\,\bigg[ 
	x_n\,\hat a_m^{(U)\dagger}\,F(\bm{c}_Q)\,t^{\bm{c}_Q}\,\frac{\bm{\mathcal{Y}}_Q}{\sqrt r}\,F(\bm{c}_Q)\,
    \frac{t^{1+\bm{c}_Q}-\epsilon^{1+2\bm{c}_Q}\,t^{-\bm{c}_Q}}{1+2\bm{c}_Q}\,\hat a_n^{(U)} \\
   &\hspace{2.3cm}\mbox{}- x_m\,\hat a_m^{(u)\dagger}\,F(\bm{c}_u)\,
	\frac{t^{1+\bm{c}_u}-\epsilon^{1+2\bm{c}_u}\,t^{-\bm{c}_u}}{1+2\bm{c}_u}\,\frac{\bm{\mathcal{Y}}_u}{\sqrt r}\,
    F(\bm{c}_u)\,t^{\bm{c}_u}\,\hat a_n^{(u)} \bigg] \,.
\end{aligned}
\end{equation}
Here $x_n=m_n/\mkk$, with $m_n$ denoting the masses of the three up-type quarks. An analogous expression holds for the couplings $g_{mn}^{(d)}$. The objects $\hat a_n^{(U)}$ and $\hat a_n^{(u)}$ are vectors in flavor space, which connect the mass eigenstates with the eigenstates in the bulk-mass basis. Using the expressions for the various objects valid in the ZMA \cite{Casagrande:2008hr}, combined with the fact that the profile function $\chi_1^{S}(t)\equiv\chi_1^{S}(\phi(t))$ peaks for values $t={\cal O}(1)$, it is straightforward to derive the scaling laws
\begin{align}
   g_{mn}^{(u)} &= {\cal O}(1)\,\frac{v_h }{ \mkk}\,F(c_{Q_m})\,F(c_{u_n}) \,, \\
	g_{mn}^{(d)} &= {\cal O}(1)\,\frac{v_h}{ \mkk}\,F(c_{Q_m})\,F(c_{d_n}) \,.
\end{align}

The above couplings display the familiar RS-GIM mechanism \cite{Agashe:2004cp}, which states that the couplings of low-lying KK resonances to light SM fermions are suppressed by an overlap factor $F(c_i)$ for each fermion, which is much smaller than 1 if the fermion is light compared with the weak scale. Importantly, however, the couplings of scalar KK resonances are {\em in addition\/} suppressed by a factor $v_h/\mkk\ll 1$. As a result, the exchange of a scalar resonance between four SM fermions gives rise to interactions proportional to 
\begin{equation}
   \frac{v_h^2}{\mkk^4}\,F(c_{Q_{n_1}})\,F(c_{q_{n_2}})\,F(c_{Q_{n_3}})\,F(c_{q_{n_4}}) \,, 
\end{equation}
which are suppressed by {\em four\/} powers of $\mkk$ and hence correspond to dimension-8 operators in the low-energy effective weak Hamiltonian. With $\mkk$ in the range of 10\,TeV, these effective interactions are highly suppressed and irrelevant for all practical purposes.  

\section{Higgs portal coupling to the bulk scalar and modified Higgs couplings}
\label{sec:Higgsportal}

An interesting possibility which we have omitted so far is to add a portal coupling $\lambda_{\Sigma H}\Sigma^2 H^\dagger H$ to the integrand of the 5D action in \eqref{eq:5Daction}, which connects the new the $\mathbb{Z}_2$-odd bulk scalar with the Higgs doublet. After electroweak symmetry breaking and the symmetry breaking giving rise to the scalar VEV $\omega(\phi)$ in \eqref{def_bulkscalar}, such an interaction induces a mixing between the Higgs field $h$ of the SM and the first scalar KK resonance $S_1$ (as well as higher KK resonances). This, in turn, has the effect of reducing the couplings of the Higgs boson to SM particles by a factor $\cos\theta_{Sh}$, where $\theta_{Sh}$ denotes the corresponding mixing angle.

\begin{figure}[t!]
\begin{center} \hspace*{-0.65cm}
\includegraphics[scale = 0.8]{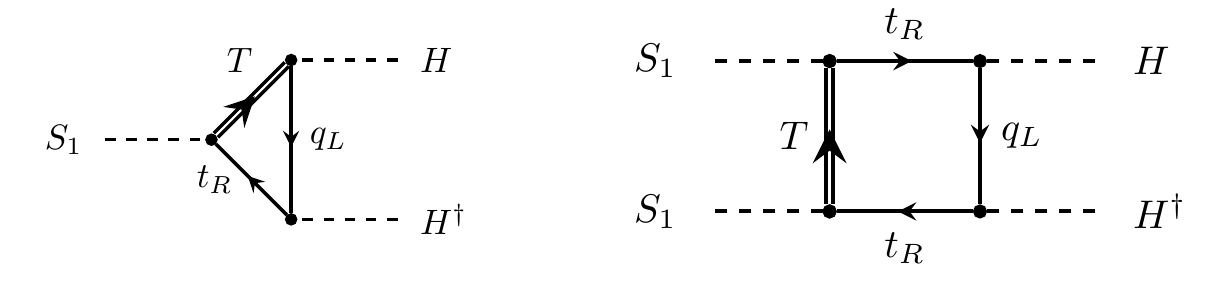} 
\vspace{2mm}
\caption{Loop diagrams leading to effective interactions coupling the first KK resonance of the bulk scalar $S_1$ to the higgs boson.}
\label{fig:Sh_diagrams}
\end{center}
\end{figure}

In the setup discussed in this paper, where the 5D Higgs field is confined to live on the IR brane, the portal coupling in the Lagrangian vanishes, since the $\mathbb{Z}_2$-odd bulk scalar field is strictly zero on the IR brane. However, such a coupling is inevitably induced by loops involving bulk fermions, which couple to the scalar field $\Sigma$ as well as to the Higgs field residing on the IR brane. Figure~\ref{fig:Sh_diagrams} shows some representative diagrams. For the mixing angle relating the physical Higgs boson $h_{\rm phys}$ to the Higgs field $h$ in the Lagrangian, we then expect
\begin{equation}
   \tan\theta_{Sh} \sim \frac{N_c}{16\pi^2}\,\frac{v_h\hat\omega}{M_S^2} 
   \sim \frac{N_c}{16\pi^2}\,\frac{v_h}{\mkk} \,.
\end{equation} 
Here the parameter $\hat\omega\sim \mkk$ is given in terms of an angular integral of the relevant profile functions appearing in the loops. 
It is naturally of order the KK mass scale, defined in~\eqref{defMKK}. The mixing angle is thus expected to be very small in this setup, due to the double suppression by a loop factor and the ratio $v_h/\mkk\lesssim 0.1$. It was therefore justified to neglect this mixing for our purposes in this paper.

We emphasize, however, that the assumption of a brane-localized Higgs sector is neither required nor particularly natural in the context of RS models. Why should the Higgs doublet be the only field confined to the boundary of the extra dimension? While originally it was thought that this is a necessary feature in order for such models to address the hierarchy problem, it is by now well known that one can construct realistic RS models in which the Higgs doublet -- like all other fields -- lives in the bulk of the extra dimension \cite{Davoudiasl:2005uu, Cacciapaglia:2006mz, Azatov:2009na, Vecchi:2010em, Archer:2012qa, Hahn:2013nza, Archer:2014jca}.\footnote{While solving the hierarchy problem, such bulk-Higgs models require a moderate fine tuning in order to keep the Higgs mass an order of magnitude below the KK mass scale.} In the context of such bulk-Higgs models, the portal interaction between the Higgs field and the $\mathbb{Z}_2$-odd scalar fields becomes a generic feature, and one expects that the mixing angle 
\begin{equation}
   \tan\theta_{Sh} \sim \frac{v_h\hat\omega}{M_S^2} \sim \frac{v_h}{\mkk} \,,
\end{equation} 
without a loop factor. In this case, the parameter $\hat\omega\sim \mkk$ is defined as
\begin{equation}
   v_h \hat\omega = \frac{\lambda_{\Sigma H}}{\sqrt{r}} \int_{0}^\pi\!d\phi\,e^{-\sigma(\phi)}\,
    \omega(\phi)\,v_{H}(\phi)\,\chi_1^{S}(\phi)\,\chi_0^{(h)}(\phi) \,,
\end{equation} 
where $v_H(\phi)$ is the profile of the Higgs VEV along the extra dimension (see e.g.\ \cite{Malm:2013jia}), and $\chi_1^{S}(\phi)$, $\chi_0^{(h)}$ are the profiles of the lowest-lying scalar and Higgs resonances, respectively. In this case the mixing angle can naturally be of order a few percent, which might bring it to the sensitivity level of future precision Higgs measurements (see e.g.\ \cite{Cepeda:2019klc, deBlas:2019rxi}).
From a phenomenological point of view, this may be the most significant imprint of this class of models. We leave a more detailed estimate of the modifications of the Higgs couplings in generalizations of our setup invoking a bulk-Higgs field for future study.

\section{Summary and conclusions}
\label{sec:conclusions}
The last decades of research in particle physics have led to a vast landscape of extensions of the SM. Motivated by the gauge hierarchy problem, scenarios like supersymmetry, composite-Higgs models and models with warped extra dimensions have played a prominent role. The current absence, so far, of any sign of new physics  at the LHC  has diversified both the motivations behind BSM physics and their experimental probes. Questions like the strong CP problem, the origin of dark matter, the flavor puzzle or the baryon asymmetry of the universe have received increasing attention, at the expense of the naturalness problem. In this regard, models of warped extra dimension still offer a beautiful interplay between several of these questions. The RS model was originally proposed to solve the gauge hierarchy problem; however, soon after its inception it was realized that it also allows for an explanation of the flavor puzzle, by means of a specific pattern of localization of the SM fermions in the bulk of the RS geometry, thus connecting the hierarchy of fermion masses to the one existing between the electroweak and Planck scales. They also allow e.g.\ for natural implementations of neutrino masses and provide extra sources of CP violation, crucial for baryogenesis.  In addition, thanks to the AdS/CFT correspondence, models with warped extra dimensions offer a precious window into strongly interacting sectors. We leave some interesting questions about the precise nature of the CFT dual of our model and the implications of the inclusion of a new scale, related to the bulk scalar VEV, in the dual theory for future work.

An important question that has not received much attention in RS models with bulk matter fields is whether the 5D fermion masses, which are responsible for the different fermion localizations along the extra dimension and therefore for solving the flavor puzzle, can be generated dynamically. These fermion bulk masses, which have to be odd under the $\mathbb{Z}_2$ orbifold symmetry of the RS geometry, are normally  assumed to be a constant parameter with an {\it ad-hoc} $\sgn(\phi)$ function. In this work, we have demonstrated that it is indeed possible to obtain these masses dynamically, via the VEV of a $\mathbb{Z}_2$-odd scalar with bulk Yukawa couplings to the different 5D fermion fields. In particular, we have shown in section~\ref{sec:back} that an odd bulk scalar can develop a non-trivial background solution both for flat and warped extra dimensions. We have shown  that kink-like solutions are only found for realistic scenarios, where the hierarchy problem is solved, and the product $kr \gg 1$. We have considered non-trivial effects of the $\phi$-dependent VEV on the background geometry in the case of a warped extra dimension. We have solved the coupled gravity-scalar Einstein equations and computed the background solutions for both the warp factor and the scalar field. 
It turns out that a non-negligible backreaction is induced on the AdS metric, which modifies the warp function in the vicinity of the UV and IR branes for reasonable values of the gravity-scalar coupling $\gamma=\abs{\mu}^2/(\lambda M_*^3)$. This is a consequence of the profile of the odd bulk scalar VEV being approximately constant in the bulk, while strongly changing in the vicinity of both branes. 
Hence, the effects of the backreaction are especially visible in those regions. Moreover, we have also briefly discussed the stability of the background geometry in the presence of an odd bulk scalar.

In section~\ref{sec:gauge}, we have analyzed the gauge sector of the minimal RS model with an $SU(2)_L\times U(1)_Y$ bulk gauge symmetry and its extension featuring a custodial symmetry in the bulk. In particular, we have calculated the oblique parameters $S$ and $T$ in the presence of a general warp function $\sigma(\phi)$ considering backreaction effects, and compared the results with those obtained in the conventional RS model (i.e.\ using the linear AdS warp function $kr\phi$). We have shown that presently allowed values of the $S$ and $T$ parameters set a stringent constraint on the masses of the first KK resonances. In particular, the mass of the first KK gluon state (which is often the lightest KK resonance in the model) has to be or order $10-14\tev$, irrespectively of whether the backreaction on the metric is included or neglected.  We have also calculated the KK spectrum of the $\mathbb{Z}_2$-odd bulk scalar and studied its main decay channels for several benchmark cases.

Our mechanism for dynamically generating the fermion bulk masses has been presented in section~\ref{sec:fermions}, where for simplicity we consider a single 5D Dirac fermion $\Psi$ which couples to the bulk odd scalar $\Sigma$ through a 5D Yukawa coupling, $ \mathcal{Y} \bar \Psi \Sigma \Psi$. After the KK decomposition of the bulk fermion, a KK tower of left-handed and right-handed chiral fermions is generated, where due to boundary conditions a zero mode exists for only one of the two chiralities. As usual, these chiral zero modes of the bulk fermion fields correspond to the SM fermions. We show how the zero-mode profile of the bulk fermions depends on the Yukawa coupling along with the $\phi$-dependent VEV of the bulk scalar. By choosing appropriate values of the bulk Yukawa couplings, the zero-mode fermions can be localized towards either one of the branes. We have calculated the deviations of the zero-mode and KK-mode profiles in our model as compared to the conventional RS case. These deviations can become sizable when the effects of the backreaction are taken into account. This, in turn, can have an impact on flavor observables and therefore on flavor bounds.

The beauty of dynamically generating bulk fermion masses is manifested when one considers the full SM fermion content and calculates the SM fermion masses after electroweak symmetry breaking. The localization of the zero-mode profiles determines these masses in the effective theory. In section~\ref{sec:pheno}, we have illustrated this mechanism for the case of a $\phi$-dependent VEV  and a modified warp factor, obtaining the hierarchical SM quark masses and mixing angles. Furthermore, we have studied the impact of the dynamical generation of bulk fermion masses on flavor observables and the RS-GIM mechanism. We have examined, in particular, the flavor observable $\epsilon_K$ in $K$--$\bar K$ mixing, which typically sets one of the most stringent bounds on models with a warped extra dimension. We have compared our predictions with those obtained in the conventional RS model. We have also studied the dependence of our results on the model parameters and the strength of the backreaction on the metric. We have shown that the current experimental value of $\epsilon_K$ puts a severe constraint on the mass of the first KK gluon state, which has to be above ${\cal O}(10)\tev$. Moreover, we have also shown that the presence of a non-negligible backreaction slightly tends to weaken the RS-GIM mechanism and, in particular, enhances the new-physics contribution to $\epsilon_K$. However, the impact is very modest overall, showing that the dynamical origin of the different fermion masses is not only an appealing possibility but also a phenomenologically viable one. In extensions of the SM with a warped extra dimension, the presence of TeV-scale KK resonances introduces new sources of flavor-changing interactions among the SM fermions. In our particular model, the exchange of scalar KK resonances between four SM fermions gives rise to effective dimension 8 interactions which are suppressed by four powers of $v/\mkk$ and are therefore negligible.

Finally, in section \ref{sec:Higgsportal} we briefly comment the portal coupling of the bulk scalar to the Higgs doublet. In realistic RS models the Higgs doublet should be allowed to propagate in the bulk of the extra dimension and, in the context of such bulk-Higgs models, the portal interaction between the Higgs field and the $\mathbb{Z}_2$-odd scalar fields becomes a generic feature. In these scenarios, the couplings of the physical Higgs boson could naturally be reduced, compared with the SM, by a few percent. These effects may be testable in future precision Higgs measurements. We leave a more detailed study of the modifications of the Higgs couplings in generalizations of our setup invoking a bulk-Higgs field for future study.

The results presented in this work put RS models with bulk matter fields on firmer theoretical ground, as they offer a viable dynamical mechanism for generating the crucial 5D fermion bulk masses from the VEV of a bulk scalar field. The prediction of the existence of heavy scalar KK resonances, with masses several times heavier than the masses of other KK states, is a generic consequence of this mechanism. These scalar resonances are too heavy to be produced even in a possible high-energy extension of the LHC, and as we have shown their quantum effects as virtual particles are likely to be too small to be observable. However, it may be promising to search for such particles at a future 100\,TeV collider. Their discovery would be a strong hint that the solution to the flavor puzzle offered by RS models is correct.


\section*{Acknowledgements}
This work has been supported by the Cluster of Excellence ``Precision Physics, Fundamental Interactions, and Structure of Matter'' (PRISMA$^+$ EXC 2118/1) funded by the German Research Foundation (DFG) within the German Excellence Strategy (Project ID 39083149), and by grants 05H12UME and 05H18UMCA1 of the German Federal Ministry for Education and Research (BMBF). The research of A.A.\ was also supported by FWO under the EOS--be.h project no.\ 30820817, while the work of J.C.R. was also funded by the DFG Research Training Group Symmetry Breaking in Fundamental Interactions (GRK 1581). 

\pagebreak

\appendix

\section{Electroweak precision observables in the custodial model}
\label{app:gauge}
Here we expand on the discussion from section~\ref{sec:gauge}, where we have studied the impact of the deviations from the original RS metric (caused by the backreaction) on EWPT and present general expressions for the gauge boson masses, the Fermi constant and the electroweak observables. Since the minimal case has been presented in greater detail on section~\ref{sec:gauge} and analogous expressions for this case can be found in the literature (see e.g.\ \cite{Cabrer:2011fb}), we focus here on the extended model with a custodial symmetry in the bulk. In this case the gauge-sector bulk action reads \cite{Casagrande:2010si}
\beq
\begin{split}
S^{\rm gauge}_{\rm 5D} = \, & \int d^4x\int_{-\pi}^\pi d\phi\, \sqrt{g}\,g^{KM} g^{LN}\Big[ - \frac14\,L_{KL}^a L_{MN}^a - \frac14\,R_{KL}^a R_{MN}^a -\frac14\,X_{KL}X_{MN}  \Big]\\
& +S_{\rm Higgs}  + S_{\rm GF} + S_{\rm FP} \,,	\label{gauge_cus_action}
\end{split}
\eeq
where $L_{MN},R_{MN}$, and $X_{MN}$ are the field-strength tensors for the bulk gauge groups $SU(2)_L, SU(2)_R$, and $U(1)_X$ respectively, and $a=1,2,3$.
Employing the usual redefinition of the Higgs field to canonically normalize its kinetic term, the action for the Higgs sector $S_{\rm Higgs}$ takes the form
\beq
 S_{\rm Higgs} = \int d^4x \int_{-\pi}^{\pi} d\phi \Big[\frac{1}{2}\mathrm{Tr}\left[(D_\mu H)^\dagger\,(D^\mu H)\right] - V( H) \Big] \delta(\phi-\pi) \,,
\eeq
where the Higgs bi-doublet can be represented as
\beq    
H(x) = \frac{1}{\sqrt2}\begin{pmatrix} \; v_h+h(x)-iG^3(x)&-i\sqrt2\,G^+(x) \\
-i\sqrt{2}G^-(x) & v_{h} + h(x) + iG^3(x) \;\;
\end{pmatrix} \, .
\eeq
The covariant derivative on the Higgs field is defined as 
\beq
iD_\mu H  = i\partial_\mu H+\frac{g_{5L}}{2}L_{\mu}^a \tau^a H -\frac{g_{5R}}{2}H \tau^a R_{\mu}^a \,,   \label{co_cus_dir_M_SM}
\eeq
with $g_{5L}$ and $g_{5R}$ denoting the 5D gauge couplings of $SU(2)_L$ and $SU(2)_R$, respectively.
By means of the field redefinitions
\beq
L_{\mu}^{\pm}=\frac{1}{\sqrt{2}}(L_{\mu}^1\mp iL_{\mu}^2),\qquad R_{\mu}^{\pm}=\frac{1}{\sqrt{2}}(R_{\mu}^1\mp iR_{\mu}^2)
\eeq
the equations of motion for the charged gauge bosons can be brought to the same form as for the minimal case, c.f.\ \eqref{eq:gauge_eom}. We obtain
\beq
   - \frac{1}{r^2}\,\partial_\phi\Big(e^{-2\sigma(\phi)}\,
   \partial_\phi\,\chi_n^{\mathbb{W}}(\phi) \Big)= (m_n^{\pm})^2\,\chi_n^{\mathbb{W}}(\phi) 
   \,;\qquad \mathbb{W}=L^{\pm},R^{\pm},\label{eq:cgauge_eom}
\eeq
with non-diagonal boundary conditions 
\beq    
\begin{matrix*}[r] \; \partial_\phi\,\chi_n^{L^{\pm}}(0) = 0 &  \\
\chi_n^{R^{\pm}}(0) = 0 & \;\;
\end{matrix*} , \qquad \quad
\Bigg[ \mathbb{I} \, \partial_\phi + \pi e^{2\sigma(\pi)}r^2 \tilde{m}_{\tilde{A}}^2 
\begin{pmatrix} c_W^2 & -s_W c_W \\
-s_W c_W & s_W^2
\end{pmatrix} \Bigg] \begin{pmatrix} \chi_n^{L^{\pm}}(\pi^-) \\
\chi_n^{R^{\pm}}(\pi^-)
\end{pmatrix} = 0
,
\eeq
where we have defined
\begin{align}
	\tilde{m}_{\tilde{A}}^2=\frac{1}{2\pi r}\frac{(g_{L5}^2+g_{R5}^2) v_{h}^2}{4}\,,\qquad s_W=\frac{g_{5R}}{\sqrt{g_{5L}^2+g_{5R}^2}}\,.
\end{align}
Similarly, if one defines
\begin{align}
	\begin{pmatrix} Z_{M}\\ A_M\end{pmatrix}=\begin{pmatrix} c_w& -s_w\\ s_w& c_w\end{pmatrix}\begin{pmatrix}L_{M}^3 \\ B_M \end{pmatrix},\qquad 	\begin{pmatrix} Z_{M}^{\prime}\\ B_M\end{pmatrix}=\begin{pmatrix} c_\theta& -s_\theta\\ s_\theta& c_\theta\end{pmatrix}\begin{pmatrix}R_{M}^3 \\ X_M \end{pmatrix} ,
\end{align}
with
\begin{align}
	s_\theta=\frac{g_{5X}}{\sqrt{g_{5R}^2+g_{5X}^2}}\,, \qquad s_w=\frac{g_{5Y}}{\sqrt{g_{5L}^2+g_{5Y}^2}}\,,\qquad g_{5Y}=\frac{g_{5R}g_{5X}}{\sqrt{g_{5R}^2+g_{5X}^2}}\,,
\end{align}
the bulk equations of motion for the neutral gauge fields become
\begin{align}
   - \frac{1}{r^2}\,\partial_\phi\Big(e^{-2\sigma(\phi)}\,
	\partial_\phi\,\chi_n^{\mathbb{Z}}(\phi) \Big)&= (m_n)^2\,\chi_n^{\mathbb{Z}}(\phi) 
	\,;\qquad \mathbb{Z}=Z^{\prime},Z,\\
   - \frac{1}{r^2}\,\partial_\phi\Big(e^{-2\sigma(\phi)}\,
   \partial_\phi\,\chi_n^{A}(\phi) \Big)&= (m_n^A)^2\,\chi_n^{A}(\phi) 
   \,,
\end{align}
with the boundary conditions
\beq \begin{aligned} \partial_{\phi}\,\chi_n^{A}(0) = \partial_{\phi}\,\chi_n^{A}(\pi^-)=0 \, , \qquad \partial_\phi\,\chi_n^{Z}(0) = 0 \, , \qquad \chi_n^{Z^\prime}(0) = 0 \,, \qquad \\
\Bigg[ \mathbb{I} \, \partial_\phi + \pi e^{2\sigma(\pi)}r^2 \tilde{m}_{\tilde{A}}^2 
\begin{pmatrix} c_W^2 / c_w^2 & - s_W c_W c_\theta / c_w \\
- s_W c_W c_\theta / c_w & s_W^2 c_\theta^2
\end{pmatrix} \Bigg] \begin{pmatrix} \chi_n^{Z}(\pi^-) \\
\chi_n^{Z^\prime}(\pi^-)
\end{pmatrix} = 0
\,. \end{aligned}
\eeq
We can solve the equations of motion for the zero modes of the charged and neutral gauge bosons perturbatively in powers of $v_h^2/\mkk^2$. 
For the masses of the zero-mode gauge bosons, we then obtain
\begin{align}\label{A.14}
	m_W^2 = \tilde{m}_{\tilde{A}}^2c_W^2 \Bigg[ 1 - & \dfrac{r^2 \tilde{m}_{\tilde{A}}^2}{\pi} \int_0^{\pi} d\phi_1 e^{2\sigma(\phi_1)} \left( c_W^2\phi_1^2 + \pi^2 s_W^2 \right) \Bigg] +\mathcal{O}\left(\frac{v_h^6}{\mkk^4}\right) ,\\
	m_Z^2 = \tilde{m}_{\tilde{A}}^2 \dfrac{c_W^2}{c_w^2} \Bigg[ 1 - & \dfrac{r^2 \tilde{m}_{\tilde{A}}^2}{\pi} \int_0^{\pi} d\phi_1 e^{2\sigma(\phi_1)} \left(\dfrac{c_W^2}{c_w^2}\phi_1^2 + \pi^2 s_W^2 c_\theta^2 \right) \Bigg] +\mathcal{O}\left(\frac{v_h^6}{\mkk^4}\right) ,
\end{align}
while the photon remains massless and has a constant profile in the bulk. Moreover, we find for the Fermi constant
\begin{equation}\label{A.15}
\begin{split}
	\frac{G_F}{\sqrt{2}}&=\frac{g_{5L}^2}{8rm_W^2}\left[(\chi_0^{L\pm}(0))^2-m_W^2\tilde{D}_{W}(0,0;0)\right] \\
	& = \frac{1}{2v_h^2} \Bigg[ 1 + r^2 \tilde{m}_{\tilde{A}}^2 \pi \int_0^\pi d\phi_1e^{2\sigma(\phi_1)} \left( 1 + \rho_W \right) +\mathcal{O}\left(\frac{v^4_h}{\mkk^4} \right) \Bigg].
\end{split}
\end{equation}
The oblique parameters $\hat{S}$, $\hat{T}$, $\hat{W}$ and $\hat{Y}$ defined in \cite{Barbieri:2004qk} can be expressed in terms of the quantities \cite{Davoudiasl:2009cd}
\begin{align}
	\hat{\alpha} & = 2\pi\left[\tilde{D}_{\mathbb{A}}(\pi,\pi;0)-D^{(-)}(\pi,\pi;0)\right] = \pi r^2\int_0^{\pi}d\phi_1 \, e^{2\sigma(\phi_1)} \left(1 - \frac{\phi_1^2}{\pi^2} \right) , \\
	\hat{\beta} & = 2\pi \tilde{D}_{\mathbb{A}}(0,\pi;0) = \pi r^2 \int_0^{\pi}d\phi_1 \, e^{2\sigma(\phi_1)} \frac{\phi_1}{\pi}\left( 1 - \frac{\phi_1}{\pi}\right) , \\
	\hat{\gamma} & = 2\pi \tilde{D}_{\mathbb{A}}(0,0;0) = - \pi r^2\int_0^\pi d\phi_1 e^{2\sigma(\phi_1)} \left( 1 - \frac{\phi_1}{\pi} \right)^2 ,
\end{align}
where $\tilde{D}_{\mathbb{A}}(\phi,\phi^{\prime};\hat{p})$ was defined in \eqref{eq:5dprop}, and $D^{(-)}(\phi,\phi^{\prime};0)$ is the 5D propagator for bulk gauge fields with UV Dirichlet boundary conditions, evaluated at zero momentum and before electroweak symmetry breaking. One finds 
\begin{equation}
	D^{(-)}(\phi,\phi^{\prime};0) = -\frac{r^2}{2} \int_0^{\phi_<} d \phi_1\, e^{2\sigma(\phi_1)} \,,
\end{equation}
where $\phi_{<}=\min(\phi,\phi^{\prime})$. This leads to  
\begin{align}
	\hat{S} & = \frac{g^2v_{h}^2}{2}\big(\hat{\beta}-\hat{\gamma}\big) 
	= \frac{2 \alpha \pi v_h^2}{s_w^2} \,\pi r^2 \int_0^{\pi} d\phi_1 \, e^{2\sigma(\phi_1)}\left( 1 - \frac{\phi_1}{\pi} \right) ,\\
	\hat{T} & = \frac{g^{\prime 2}v_{h}^2}{4}\left(-\hat{\alpha}+2\hat{\beta}-\hat{\gamma}\right)=0,\\
	\hat{W} & = \hat{Y} = -\frac{g^2 v_{h}^2}{4}\hat{\gamma}
	= \frac{\alpha  v_h^2}{s_w^2 }\,\pi^2 r^2 \int_0^{\pi}d\phi_1e^{2\sigma(\phi_1)} \left( 1 - \frac{\phi_1}{\pi} \right)^2 ,
\end{align}
where $g\equiv g_{5L}/\sqrt{2\pi r}$ and $g^{\prime}\equiv g_{5Y}/\sqrt{2\pi r}$ are the electroweak gauge couplings of the SM.

In the limit of negligible backreaction on the metric, these low-energy observables in \eqref{A.14} and \eqref{A.15} take the familiar form
\begin{align}
	m_W^2&=c_W^2\tilde{m}_{\tilde{A}}^2\left[1-\frac{c_W^2\tilde{m}_{\tilde{A}}^2}{2\mkk^2}\left(L-1+\frac{1}{2L}\right)-\frac{s_W^2\tilde{m}_{\tilde{A}}^2}{2 \mkk^2}L+\mathcal{O}\left(\frac{v_{h}^4}{\mkk^4}\right)\right],\\
	m_Z^2&=\frac{c_W^2}{c_w^2}\tilde{m}_{\tilde{A}}^2\left[1-\frac{c_W^2\tilde{m}_{\tilde{A}}^2}{2c_w^2\mkk^2}\left(L-1+\frac{1}{2L}\right)-\frac{s_W^2c_{\theta}^2\tilde{m}_{\tilde{A}}^2}{2 \mkk^2}L+\mathcal{O}\left(\frac{v_{h}^4}{\mkk^4}\right)\right],\\
	\frac{G_F}{\sqrt{2}}&=\frac{1}{2v^2_h}\left[1+\frac{\tilde{m}_{\tilde{A}}^2}{2 \mkk^2}L+\mathcal{O}\left(\frac{v_{h}^4}{\mkk^4}\right)\right]=\frac{1}{2v_{\rm SM}^2},
\end{align}
where $L\equiv kr\pi$. On the other hand, the expressions for the oblique parameters $S$ and $T$ simplify to
\begin{equation}
	S = \frac{4s_w^2}{\alpha}\,\hat{S} = \frac{2\pi v_{h}^2}{\mkk^2} \,, \qquad 
	T = \frac{\hat{T}}{\alpha} = 0 \,.
\end{equation}



\bibliography{bib_warped}{}
\bibliographystyle{aabib}

\end{document}